\documentclass[pra,superscriptaddress,showpacs,papersize=a4paper,twocolumn,floatfix]{revtex4-1}

\usepackage[pdftex]{graphicx}
\usepackage{bm}
\usepackage{etoolbox}
\usepackage[usenames,dvipsnames]{color}
\usepackage{mathtools}
\usepackage{amsmath}
\usepackage{amsfonts}
\usepackage{amssymb}

\DeclarePairedDelimiter\bra{\langle}{\rvert}
\DeclarePairedDelimiter\ket{\lvert}{\rangle}
\DeclarePairedDelimiterX\braket[2]{\langle}{\rangle}{#1 \delimsize\vert #2}

\newcommand{\bg}{ \begin{gather} }
\newcommand{\eg}{\end{gather}}
\newcommand{\be}{ \begin{equation} }
\newcommand{\ee}{\end{equation}}
\newcommand{\bea}{ \begin{eqnarray} }
\newcommand{\eea}{\end{eqnarray}}

\def\Tr{\mathop{\rm Tr}}

\newcommand{\str}{\mathop{\rm Str}}

\renewcommand{\Re}{\mathop{\rm Re}}
\renewcommand{\Im}{\mathop{\rm Im}}

\AtEndEnvironment{thebibliography}{}

\begin{document}

\title{Statistics of eigenstates near the localization transition on random regular graphs}

\author{K.\,S.~Tikhonov}
\affiliation{Institut f{\"u}r Nanotechnologie, Karlsruhe Institute of Technology, 76021 Karlsruhe, Germany}
\affiliation{Condensed-matter Physics Laboratory, National Research University Higher  School of Economics, 101000 Moscow, Russia}
\affiliation{L.\,D.~Landau Institute for Theoretical Physics RAS, 119334 Moscow, Russia}

\author{A.\,D.~Mirlin}
\affiliation{Institut f{\"u}r Nanotechnologie, Karlsruhe Institute of Technology, 76021 Karlsruhe, Germany}
\affiliation{Institut f{\"u}r Theorie der Kondensierten Materie, Karlsruhe Institute of Technology, 76128 Karlsruhe, Germany}
\affiliation{L.\,D.~Landau Institute for Theoretical Physics RAS, 119334 Moscow, Russia}
\affiliation{Petersburg Nuclear Physics Institute,188300 St.\,Petersburg, Russia.}

\begin{abstract}
Dynamical and spatial correlations of eigenfunctions as well as energy level correlations in the Anderson model on random regular graphs (RRG) are studied. We consider the critical point of the Anderson transition and the delocalized phase. In the delocalized phase near the transition point, the observables show a broad critical regime for system sizes $N$ below the correlation volume $N_{\xi}$ and then cross over to the ergodic  behavior.  Eigenstate correlations allow us to visualize the correlation length $\xi \sim \log N_{\xi}$ that controls the finite-size scaling near the transition. The critical-to-ergodic crossover is very peculiar, since the critical point is similar to the localized phase, whereas the ergodic regime is characterized by very fast ``diffusion'' which is similar to the ballistic transport. In particular, the return probability crosses over from a logarithmically slow variation with time in the critical regime to an exponentially fast decay in the ergodic regime. Spectral correlations in the delocalized phase near the transition are characterized by level number variance $\Sigma_2(\omega)$ crossing over, with increasing freqyency $\omega$, from ergodic behavior $\Sigma_2=\left(2/\pi^2\right)\ln\omega/\Delta$ to $\Sigma_2\propto \omega^2$ at $\omega_c\sim (N N_{\xi})^{-1/2}$ and finally to Poissonian behavior $\Sigma_2 = \omega/\Delta$ at $\omega_{\xi}\sim N_{\xi}^{-1}$. We find a perfect agreement between results of exact diagonalization and those resulting from the solution of the self-consistency equation obtained within the saddle-point analysis of the effective supersymmetric action. We show that the RRG model can be viewed as an intricate $d\to\infty$ limit of the Anderson model in $d$ spatial dimensions.
\end{abstract}
\maketitle


\section{Introduction}
\label{s1}

Anderson localization\cite{anderson58} and, in particular, transitions between localized and delocalized phases\cite{evers08} belong to central themes of the condensed matter physics. In the conventional formulation, the Anderson localization refers to a problem of non-interacting quantum particles subjected to a random potential in $d$ dimensions. Statistical properties of single-particle eigenfunctions and energy levels in a finite system at Anderson-transition critical points as well as in delocalized and localized regimes have been extensively studied, see Refs.~\onlinecite{mirlin00,evers08} for reviews. 

In recent years, the problem of many-body localization (MBL)  in interacting disordered systems at non-zero temperature\cite{gornyi2005interacting,basko2006metal} has attracted a great deal of attention.  This interest is motivated by relevance of the MBL problem to low-temperature transport properties of a very broad class of quantum systems as well as by its close connection to a very general question of ergodicity in complex systems. 
The prediction of the MBL transition\cite{gornyi2005interacting,basko2006metal} has been supported by a number of subsequent analytical and numerical works, see, in particular, Refs.~\onlinecite{oganesyan07,monthus10,kjall14,gopalakrishnan14,luitz15,nandkishore15,karrasch15,imbrie16,imbrie16a,gornyi-adp16}. Experimentally, the MBL transition and the associated physics was studied in systems of cold atoms and ions in optical traps\cite{schreiber2015observation,choi2016exploring,kondov2015disorder,smith2016many,bordia15,lueschen16,Zhang17}, spin defects in a solid state\cite{Choi17a,Choi16,choi2017observation,Ho17,choi2017observation}, and superconducting qubits\cite{roushan17}, as well as in InO films~\cite{ovadyahu1,ovadyahu2,ovadia2015evidence}.

It was proposed in Ref.~\onlinecite{altshuler1997quasiparticle} in the context of the analysis of quasiparticle decay in a quantum dot that the many-body physics can be approximated by a model of a single quantum particle moving on a Bethe lattice. Extensions of this idea of a relation between the quantum many-body dynamics and the single-particle physics on tree-like graphs have been very useful for analytical studies of MBL problems. The tree-like graphs with fixed coordination number (and without boundary) are known as random regular graphs (RRG). As was shown recently\cite{tikhonov18}, there is a particularly close connection between 
the Anderson model on RRG and the MBL transition in a model with long-range interaction (decaying as a power-law of distance). The many-body delocalization by power-law interaction has been earlier studied in a number of theoretical works\cite{burin06,Demler14,Burin15,Gutman16}. Experimentally, the long-range-interaction model is relevant for a variety of systems, including localized electron states, spin defects, superconducting circuits, trapped ions, and others. 

A close relative of the Anderson model on RRG is the sparse random matrix (SRM) ensemble (also known as Erd\"os-R\'enyi graphs in mathematical literature) studied analytically in Refs.~\onlinecite{mirlin1991universality,fyodorov1991localization,fyodorov1992novel}. The difference between the two models is that the coordination number is strictly fixed in RRG but can fluctuate in SRM.  These fluctuations are not important for the localization-related physics. 
The key property of the RRG and SRM ensembles is that they represent tree-like models without boundary (and with loops of typical size $\sim \ln N$, where $N$ is the number of sites). 

It was shown in Refs.~\onlinecite{mirlin1991universality,fyodorov1991localization,fyodorov1992novel} that the delocalized phase on the infinite cluster of the SRM model has ergodic properties in the large-$N$ limit. More specifically, it was found that (i) the levels statistics has the Wigner-Dyson form and (ii) the inverse participation ratio (IPR) $P_2=\sum_i|\psi_i|^4$, with  $\psi_i$ being the amplitude of a wave function $\psi$  on site $i$, scales with the system volume as $1/N$. 
The derivation of these results was based on a certain functional-integral representation of the correlation functions of the model in the framework of the supersymmetry formalism. In the large-$N$ limit, the integral can be evaluated by the saddle-point method. The corresponding saddle-point equation has a form analogous to the self-consistency equations obtained for the Anderson model\cite{abou1973selfconsistent,mirlin1991localization}
and the $\sigma$ model\cite{efetov1985anderson,zirnbauer1986localization,zirnbauer1986anderson,efetov1987density,efetov1987anderson,verbaarschot1988graded} on an infinite Bethe lattice. On the delocalized side of the transition, the symmetry is spontaneously broken, which results in a manifold of saddle points\cite{mirlin1991universality,fyodorov1991localization,fyodorov1992novel}. Integration over this manifold yields ergodic properties of the level and eigenfunctions statistics quoted above. 

More recently, there was a resurgence of interest to the Anderson models on RRG and on related tree-like graphs, largely in view of their relation to the MBL problems.  The works Refs.~\onlinecite{biroli12,deluca14}, which addressed the problem numerically, questioned the ergodicity of the delocalized phase in the RRG model as defined by large-$N$ limit of the energy-level statistics and of the IPR scaling.  Specifically, Ref.~\onlinecite{biroli12} found that a part of the delocalized phase is non-ergodic, while the authors of Ref.~\onlinecite{deluca14} came to the conclusion that the whole delocalized phase is non-ergodic. 
 These papers stimulated an intensive numerical research on properties of the delocalized phase in the Anderson model on RRG and related graphs \cite{tikhonov2016anderson,garcia-mata17,metz17}.  A thorough numerical study of the gap ratio characterizing the level statistics and of the IPR scaling on the delocalized side of the Anderson transition on RRG performed in Ref.~\onlinecite{tikhonov2016anderson} confirmed the analytical predictions of Refs.~\onlinecite{mirlin1991universality,fyodorov1991localization,fyodorov1992novel}. Specifically, the analysis of Ref.~\onlinecite{tikhonov2016anderson} reveals a crossover from relatively small ($N\ll N_\xi$) to large ($N\gg N_\xi$) systems, where  $N_\xi$ is the correlation volume. The values of $N_\xi$ obtained from numerical simulations agree well with the analytical prediction $\ln N_\xi \sim (W_c-W)^{-1/2}$ that
implies an exponential divergence of the correlation volume at the transition point.  For $N\ll N_\xi$ the system exhibits  a flow towards the Anderson-transition fixed point which has on RRG properties very similar to the localized phase. When the system volume $N$ exceeds $N_\xi$, the direction of flow is reversed and the system approaches its $N\to\infty$ ergodic behavior. The overall evolution with $N$ is thus non-monotonic. In combination with exponentially large values of the correlation volume $N_\xi$, this makes the finite-size analysis very non-trivial. The main conclusions of Ref.~\onlinecite{tikhonov2016anderson} have been supported by subsequent studies of the IPR scaling in the SRM-like model\cite{garcia-mata17} and of the level number variance in the RRG model\cite{metz17}. 

Before we turn to the goals of the present work, it is worth emphasizing the following important point.  When discussing the eigenstate ergodicity and its manifestations in the energy level and eigenfunction statistics, one should distinguish the RRG and similar models (that are locally tree-like but do not have boundary) from models on a finite Bethe lattice (which is a tree and thus has a boundary). A dramatic difference between two types of models  was demonstrated in Refs. \onlinecite{tikhonov2016fractality,sonner17}. Contrary to RRG model, for which delocalized states are ergodic, eigenstates on a finite Bethe lattice are characterized by multifractality with exponents that depend continuously on the position on the lattice and on the disorder strength. 
In this paper, we focus entirely on the RRG model. 

The central goal of this work is to explore systematically correlations between eigenstates of the Anderson model on RRG, at criticality and on the delocalized side of the transition. More specifically:
\begin{itemize}
\item[(i)] We study spatial correlations between amplitudes of one and the same eigenfunction. We will show that this correlation function permits to visualize directly the correlation length $\xi \sim \ln N_\xi$. 
\item[(ii)] We explore dynamical correlations, i.e. those between eigenstates separated by an energy difference $\omega$, at the same and at different spatial points.
\item[(iii)]  As a closely related dynamical observable, we analyze the time dependence of the return probability to a given spatial point. 
\item[(iv)]  We study the level number variance that characterizes the energy level statistics in the whole range of frequencies. 
\end{itemize}
This set of correlation functions is of central importance for understanding the physics of the delocalized phase and of the critical regime on RRG. 

To explore these correlations functions for large matrix sizes $N$, we use two complementary approaches. First, we perform exact diagonalization of the Anderson model on RRG. Second, we use the saddle-point approximation mentioned above, in combination with analytical and numerical solution of the resulting saddle-point equation. We find a perfect agreement between the two approaches, which serves as an additional demonstration of validity of the saddle-point solution of the problem. This is obviously an important conclusion---in particular, since the derivation of the ergodicity of the delocalized phase is based on the saddle-point analysis. 

The research program that we pursue in this paper is largely parallel to the one that has been earlier implemented for systems of finite dimensionality $d$, with a particular focus on $d\le 4$, see Refs.~\onlinecite{mirlin00,evers08} and references therein. It was argued long ago\cite{mirlin94a,mirlin94b} that the Anderson transition on tree-like graphs can be viewed as $d\to\infty$ limit of the ``conventional'' Anderson transition and that this limit is very peculiar. This peculiarity manifests itself, in particular, in the scaling of the correlation volume $N_\xi$ that determines the properties of the order-parameter function. Specifically, $N_\xi$ scales as a power law of the correlation length for any finite $d$ but exponentially on tree-like graphs. One of fundamentally important  implications of this fact is the infinite value of the upper critical dimension for the Anderson localization transition\cite{mirlin94a,mirlin94b}. In the present work we use the connection to the $d\to\infty$ limit to better understand the physics of eigenstate and energy level correlations in the Anderson model on RRG. We find that the results on RRG indeed may be viewed as matching those at large $d$ but with a highly singular character of the $d\to \infty$ limit.

The structure of the article is as follows. In Section \ref{s2} we define the microscopic model to be studied (Anderson model on RRG) and also remind the reader about the equivalent supersymmetric field-theoretical model that is obtained after the disorder averaging. 
In Section \ref{s3} we explore spatial and dynamical correlations between eigenstates as well as the time-dependent return probability.
In Section \ref{s4} an analysis of energy level correlations is performed.   Our findings are summarized in Sec.~\ref{s5}.

\section{Model}
\label{s2}
We study non-interacting spinless fermions hopping over a random regular graph (RRG) with connectivity $p = m+1$  in a potential disorder,
\begin{equation}
\label{H}
\mathcal{H}=t\sum_{\left<i, j\right>}\left(c_i^\dagger c_j + c_j^\dagger c_i\right)+\sum_{i=1} \epsilon_i c_i^\dagger c_i\,,
\end{equation}
where the sum is over the nearest-neighbor sites of the RRG. The energies $\epsilon_i$ are independent random variables sampled from a uniform distribution on $[-W/2,W/2]$; the hopping amplitude $t$ will be set to unity. The statistical properties of observables (such as disorder-averaged products of Green functions) in this model can be expressed in terms of certain functional integrals, either in supersymmetric\cite{mirlin1991localization,mirlin1991universality} or in the replicated version\cite{PhysRevB.37.3557,PhysRevE.90.052109,PhysRevLett.117.104101}. These approaches are largely equivalent (see, for example, Ref. \onlinecite{gruzberg96}); we will use the supersymmetric formulation in what follows.

The model, defined by Eq. (\ref{H}), has two sources of disorder: configurational one (randomness in the structure of the underlying graph) and fluctuations of on-site energies $\epsilon_i$. With structure of the graph being fixed, averaging over $\epsilon_i$ can be performed exactly.
Various disorder-averaged properties of a disordered system can be expressed in terms of averaged products of Green functions.
 If one wants to average a single Green function (like for the calculation of the averaged density of states), one introduces a four-component supervector attached to each node $i$ of the graph, 
\be
\Phi_i=(S_i^{(1)},S_i^{(2)},\chi_i,\chi_i^*)^T,\;\Phi^\dagger_i = (S_i^{(1)},S_i^{(2)},\chi_i^*,-\chi_i),
\label{4-supervector}
\ee
 where $S$ stays for real commuting and $\chi$ for anticommuting variables. 
 For the analysis of localization-related properties, averaged products of retarded and advanced Green functions should be considered. This requires doubling of the size of the supervector,  $\Phi_i = (\Phi_{i,1}, \: \Phi_{i,2})$. Here $\Phi_{i,1}$ corresponds to the retarded and $\Phi_{i,2}$ to the advanced subspace; each of them is a 4-supervector with a structure given by Eq.~(\ref{4-supervector}).
 
In general, averaged products of retarded and advanced Green functions (with energies $E+\omega/2$ and  $E-\omega/2$, respectively) can be evaluated as  superintegrals of the form\cite{mirlin1991localization}
\be
\label{superint}
 \int\prod_k[d\Phi_k] e^{-\mathcal{L}(\Phi)}U(\Phi),
\ee
where the preexponential factor $U(\Phi)$ represents the quantity in question and 
$$[d\Phi_k]=dS_{k,1}^{(1)}dS_{k,1}^{(2)}d\chi_{k,1}^*d\chi_{k,1} dS_{k,2}^{(1)}dS_{k,2}^{(2)}d\chi_{k,2}^*d\chi_{k,2}$$ 
is the supervector integration measure. The action $\mathcal{L}(\Phi)$ is given by
\bea
\label{superaction}
\displaystyle e^{-\mathcal{L}(\Phi)} &=& \int \prod_i d\epsilon_i\gamma(\epsilon_i) e^{\frac{i}{2}\Phi_i^\dagger\hat{\Lambda}(E-\epsilon_i)\Phi_i + \frac{i\omega}{4}\Phi_i^\dagger\Phi_i}
\nonumber \\
\displaystyle &\times & \prod_{\left<i,j\right>}e^{-i\Phi_i^\dagger\Phi_j},
\eea
where $\hat{\Lambda}$ is a diagonal supermatrix with the first four components (retarded sector) equal to $+1$ and the last four components (advanced sector) equal to $-1$.  Further, $\gamma(\epsilon)$ is the distribution of on-site random potentials; for the box distribution $\gamma(\epsilon)=\frac{1}{W}\theta(\frac{W}{2}-\left|\epsilon\right|)$. In the last factor in Eq.~(\ref{superaction}), the product goes over pairs $\left<i,j\right>$ of nearest neighbor sites. 

It is useful to consider also an $n$-orbital generalization of the problem (with $n\gg 1$), which can be viewed as describing an electron hoping between metallic granules located at the nodes of the same RRG. The Hamiltonian of such a granular system reads
\begin{eqnarray}
\mathcal{H} &=&
\sum_{\left<i, j\right>}\sum_{p,q=1}^n \left(t_{ij}^{pq} c_{ip}^\dagger c_{jq}  + {\rm h.c.} \right) 
\nonumber \\ 
&+&
\sum_{i}\sum_{p,q=1}^n (\epsilon_i^{pq} c_{ip}^\dagger c_{iq} + {\rm h.c.}). \label{HG}
\end{eqnarray}
For large $n$, the $n$-orbital problem can be mapped onto a supersymmetric $\sigma$ model\cite{efetov1985anderson,zirnbauer1986anderson,zirnbauer1986localization,efetov1987density,verbaarschot1988graded}. 
While the derivation of the $\sigma$ model becomes simpler under the assumption that $t_{ij}^{pq}$ and  $\epsilon_i^{pq}$ are Gaussian-distributed random variables, the mapping applies under much more general conditions. Physically, the underlying condition is the ergodicity on the scale of a single granule.
The action of the $\sigma$ model reads
\be
\label{action}
S[Q]= - J\sum_{\left<i, j\right>}\str(Q_i-Q_j)^2 + \frac{\pi\eta}{2\delta_0}\sum_i \str(\hat \Lambda Q_i).
\ee
Here $Q_i$ are $8 \times 8$ supermatrices satisfying the condition $Q^2=1$, the symbol $\str$ denotes the supertrace (defined as trace of the boson-boson block minus trace of the fermi-fermi block), and $\delta_0 = \nu^{-1} = W/n$ is the mean level-spacing on a granule. 
The dimensionless coupling constant $J$ is given by $J=\left(t/\delta_0\right)^2$, where $t$ is the characteristic amplitude of the hopping, $t^2 = \langle |t_{ij}^{pq}|^2\rangle$.
If all amplitudes are real, the microscopic model (\ref{HG}) belongs to the orthogonal (AI) symmetry class, determining the corresponding symmetry of the  $\sigma$ model. When the time-reversal symmetry is broken (e.g., the hopping amplitudes $t_{ij}^{pq}$ are complex with random phases), the symmetry class becomes unitary (A). The physics that we discuss in this paper is essentially the same in both cases. Since the unitary-symmetry case is somewhat simpler technically, we will focus on it below for the sake of transparency of exposition. In this case, $Q_i$ in Eq.~(\ref{action}) become $4\times 4$ supermatrices, and the action  (\ref{action}) acquires an additional overall factor of two.

While the $n=1$ Anderson model and its $n\gg 1$ generalization ($\sigma$ model) turn out to exhibit the same gross features, analytical calculations are usually somewhat simpler within the $\sigma$ model. The investigation of the model with $n=1$ on an infinite Bethe lattice was pioneered by Abou-Chacra et al, Ref.~\onlinecite{abou1973selfconsistent}; its solution in the framework of supersymmetry approach was obtained in Ref.~\onlinecite{mirlin1991localization}. 
The $\sigma$ model (i.e. the large-$n$ model) on an infinite Bethe lattice was solved via the supersymmetry approach in Refs.~\cite{efetov1985anderson,zirnbauer1986anderson,zirnbauer1986localization,efetov1987density,verbaarschot1988graded}. It is worth stressing that observables that can be defined in the delocalized phase on an infinite Bethe lattice are finite-frequency correlation functions. This corresponds to taking the limit $N \to \infty$ at fixed non-zero frequency (energy separation) $\omega$. This limit is exactly opposite to what one is interested in when studying correlations in a single eigenstate ($\omega=0$) or between nearby-in-energy eigenstates (small $\omega$) on a finite lattice. In the latter case, the boundary conditions are crucially important, as has been already emphasized in Sec.~\ref{s1}:  the RRG studied in this paper is essentially different from a finite Bethe lattice. 

Our analytical treatment of the $n=1$ model defined by Eq. (\ref{H}) will be based on the supersymmetry approach, Eqs. (\ref{superint}) and (\ref{superaction}). 
Upon averaging over configurational disorder, one obtains an action that can be treated, in the limit of a large system ($N\gg 1$) via saddle-point approximation.
As discussed in more detail in Sec. \ref{s3}, the resulting saddle-point equation is equivalent to the the self-consistency equation for the same ($n=1$) model defined on an infinite Bethe lattice. This is a very important property which is manifestation of the fact that, with probability unity, RRG has locally a structure of a tree with fixed connectivity $p$ in the vicinity of any of its sites. [The mathematically accurate statement is that, at large $N$,   a $1-o(1)$  portion of RRG nodes  (i.e., almost all nodes) have their $\kappa\log_m N$-neighbourhood loopless for $\kappa<1/2$.]  In a similar way, an $n$-orbital model (with $n\gg 1$) on RRG, Eq.~(\ref{HG}), is described by the $\sigma$ model, Eq.~(\ref{action}), which leads, in the limit $N\to\infty$ to the corresponding Bethe-lattice self-consistency equation. We emphasize once more the crucial difference between RRG and a finite Bethe lattice: investigation of the eigenfunction statistics in the latter case leads to a recurrence (rather than self-consistency) relation, see Refs. \onlinecite{tikhonov2016fractality,sonner17}, resulting in multifractality of wave functions.

\section{Wavefunction correlations}
\label{s3}

\subsection{From finite-$d$ models to RRG}
\label{s3.1}

One of the most remarkable features of Anderson transitions in $d$ dimensions is multifractal spatial statistics of critical wavefunctions. We begin this section by reminding the reader of the basic results on eigenfunction statistics for finite $d$; after this we will return to the case of RRG.

We consider first the scaling of the average inverse participation ratio (IPR) of eigenstates, $P_2=\left<\int d^d r|\psi(r)|^{4}\right>$, with the (linear) system size $L$ in the large-$L$ limit. Here angular brackets denote averaging over the ensemble of random Hamiltonians. It is well known that three different scaling laws exist depending on the phase of the system\cite{evers08}:
\be
\label{P2sc}
P_2 \propto
\left\{
\begin{array}{ll}
 L^{-d}, & \quad \textrm{metallic;}\\
 L^{-d-\Delta_2}, & \quad \textrm{critical;}\\
L^0, & \quad \textrm{localized.}
\end{array} 
\right.
\ee 
The second line of Eq.~(\ref{P2sc}) describes the fractal scaling at the critical point of the Anderson transition characterized by an anomalous dimension $\Delta_2$ (which depends on the spatial dimensionality $d$ and satisfies $-d < \Delta_2 <0$). 

More generally, one can study correlations of the same wavefunction at different spatial points defined formally as 
\be
\label{alphadef}
\alpha_E \left( r_{1},r_{2}\right) =\Delta \left\langle
\sum_{k}\left\vert \psi _{k}\left( r_{1}\right) \right\vert ^{2}\left\vert
\psi _{k}\left( r_{2}\right) \right\vert ^{2}\delta \left( E-E_{k}\right) \right\rangle.
\ee
Here $\psi_k$ are eigenstates and $E_{k}$ the corresponding energy levels,  $E$ is the energy at which the statistics is studied, $\Delta=1/\nu(E)N$ is the mean level spacing, and  $\nu(E)=N^{-1}\left<\Tr\delta(E-\hat H)\right>$ is the density of states.
For coinciding points, this correlation functions reduces to the IPR,  $P_2 = \int d^dr \alpha_E(r,r) \sim L^d \alpha_E(r,r)$. At finite spatial separation, a  wavefunction in the delocalized phase near the Anderson transition exhibits strong self-correlations up to the correlation length $\xi$,
\be
\label{alphad}
L^{2d}\alpha_E(r_1,r_2)\sim \left(|r_1-r_2|/\min(L,\xi)\right)^{\Delta_2},
\ee
for $|r_1-r_2| < \xi$.
At the Anderson-transition point, the correlation length $\xi$ diverges, and the critical correlations (\ref{alphad}) extend over the whole system. If the system is slightly off the Anderson-transition point ($\xi$ finite but large), it remains effectively critical as long as $L \ll \xi$. 

Further, we recall finite-$d$ properties of correlations of different (but close in energy) wavefunctions. The corresponding correlation function is formally defined as 
\bea
\label{sigmadef}
&& \beta_E \left( r_{1},r_{2},\omega \right) = \Delta
^{2}R_E^{-1}\left( \omega \right) \nonumber\\  && \times \left\langle \sum_{k\neq l}\left\vert \psi
_{k}\left( r_{1}\right) \psi _{l}\left( r_{2}\right) \right\vert ^{2}\delta
\left( E- \frac{\omega}{2}-E_{k}\right) \delta \left( E+\frac{\omega}{2} -E_{l}\right)
\right\rangle, \nonumber \\
\eea
where $\omega$ is the energy difference between the states, and we have introduced the level correlation function
\be
\label{R-omega}
R_E(\omega)=\frac{1}{\nu^2(E)}\left<\nu(E-\omega/2)\nu(E+\omega/2)\right>.
\ee
The correlation function (\ref{sigmadef}) exhibits the scaling
\be
\label{betad}
L^{2d}\beta_E(r_1,r_2,\omega)\sim \left(|r_1-r_2|/\min(L_\omega,\xi)\right)^{\Delta_2}
\ee
for $|r_1-r_2|<\min(L_\omega,\xi)$. Here $L_{\omega}\sim (\omega\nu)^{-1/d}$ is the length scale associated with the frequency $\omega$ (or, equivalently, with the time $\sim \omega^{-1}$) at criticality. The exponent $\Delta_2$ determines also the scaling of the diffusion propagator at criticality\cite{chalker1990scaling}.

One can extend the above analysis to arbitrary moments of the wavefunction, $P_q=\left<\int d^d r|\psi(r)|^{2q}\right>$ whose scaling at criticality defines a spectrum of multifractal exponents $\Delta_q$. It was shown that the multifractality can be used to devise an effecient finite-size scaling procedure\cite{PhysRevB.91.184206, rodriguez2011multifractal}. Recently, multifractal properties of the critical wavefunctions have attracted additional attention in view of their importance for interaction effects\cite{kettemann2009critical,feigel2010fractal,burmistrov2012enhancement,mayoh2015global,burmistrov2011wave,feigel2018electron,PhysRevLett.109.246801,PhysRevB.85.085122}.

After this reminder, we turn to the analysis of the eigenfunction statistics on RRG. First, we will formulate conjectures for the RRG correlation functions based on an extrapolation of the finite-$d$ results to $d \to\infty$. Then we will perform their accurate derivation in the framework of the supersymmetric field theory, which will support these conjectures and also make them more precise. Finally, we will corroborate the analytical results by exact-diagonalization numerics. Our main focus will be on system at the critical point as well as in the delocalized phase not too far from the transition. 

To understand what happens with the multifractal exponent $\Delta_2$ in the extrapolation to  RRG ($d\to\infty$), we note that
the scaling of IPR at criticality is on RRG the same as in the localized phase, $P_2\sim 1$. The fact the IPR remains of order of unity when the system approaches the critical point from the localized side is known for $n \gg 1$ 
\cite{zirnbauer1986anderson} and $n=1$ \cite{mirlin1991localization} models on an infinite Bethe lattice. Since boundary conditions do not matter for localized phase wave functions, this also applies to the localized phase on RRG. By virtue of continuity, it follows that $P_2\sim 1$ holds also at criticality on RRG.  
Comparing with Eq.~(\ref{P2sc}), we see that the RRG behavior of IPR can be interpreted as the $d\to\infty$ limit of the finite-$d$ result with $\Delta_2(d\to\infty) \to -d$, which corresponds to the strongest possible multifractality. 

 Now we translate Eq. (\ref{alphad}) to RRG. First, we use the fact that the factor $L^d$ is nothing but the system volume and thus becomes the RRG number of sites $N$.  Second, we use the limit $\Delta_2 \to -d$ for the multifractal exponent. Finally, we replace the factors of the type $r^d$, which represent a number of sites within a distance $r$ from a given site, by their RRG counterpart $m^r$.  This yields the following conjecture for the scaling of the self-correlation function (\ref{alphad}) on RRG:
\be
\label{P2sc2}
\alpha_E(r_1,r_2)\sim
\left\{
\begin{array}{ll}
 N^{-2} m^{\xi-r_{12}}, & \quad \textrm{metallic}, \ \ r_{12} < \xi;\\
N^{-1}m^{-r_{12}}, & \quad \textrm{critical.}
\end{array}
\right.
\ee
Here $r_{12}$ is the distance (the length of the shortest path) between the two points $r_1$ and $r_2$ on RRG.
Proceeding in the same with the correlation function of two different eigenfunctions, Eq.~(\ref{betad}), we obtain its expected behavior on RRG:
\be
\label{betasc}
\beta_E(r_1,r_2,\omega)\sim
\left\{
\begin{array}{ll}
 N^{-2} m^{\xi-r_{12}}, & \quad \textrm{metallic}, \ \ r_{12} < \xi;\\
N^{-2} \omega^{-1}m^{-r_{12}}, & \quad \textrm{critical.}
\end{array}
\right.
\ee
In what follows, we will derive more rigorously these scaling formulas [including also subleading factors missing in Eqs.~(\ref{P2sc2}) and (\ref{betasc})] and will compare the analytical results (combined with numerical solution of the saddle-point equation) with those of the exact diagonalization. 

\subsection{Field-theoretical approach}
\label{s3.2}

To proceed with the analytical derivation, we use the approach based on supersymmetric field theory. Such a derivation was performed, with applications to the level statistics and to the IPR scaling in the SRM model, in Refs.\cite{mirlin1991universality,fyodorov1991localization}. We will follow these works (with minor modification related to  a difference between the RRG and SRM ensembles) and generalize the results to other eigenfunction correlations.  It is worth pointing out that a closely related saddle-point analysis was performed in the replica formalism for the level statistics on RRG in Refs.~\cite{PhysRevE.90.052109,PhysRevLett.117.104101,metz2017level}. The self-consistency equations representing the saddle-pont condition, as obtained in the supersymmetry and replica approaches, are equivalent. 

In order to obtain the partition function of the theory Eqs. (\ref{superint}), (\ref{superaction}) averaged over the RRG ensemble, we consider an ensemble of $N\times N$ Hamiltonians with the following joint distribution of diagonal $H_{ii}$  and off-diagonal $H_{ij} = H_{ji} = A_{ij}t_{ij}$ matrix elements:
\bea
&& {\cal P}(\{H_{ii}\}, \{A_{ij}\}, \{t_{ij}\}) =  \prod_i \gamma (H_{ii}) \nonumber \\
&& \hspace*{1cm} \times  \prod_{i < j} \left [ \left( 1 - \frac{p}{N} \right) \delta(A_{ij}) + \frac{p}{N} \delta(A_{ij} - 1) \right] \nonumber \\
&& \hspace*{1cm} \times   \prod_i  \delta \left ( \sum_{j \ne i} A_{ij} - p \right) \prod_{i<j} h(t_{ij}) .
\label{rrg-distribution}
\eea
Here $A_{ij}$ is the adjacency matrix. For the purpose of generality (and for simplifying a comparison with Refs.~\cite{mirlin1991universality,fyodorov1991localization}), we have included in Eq.~(\ref{rrg-distribution}) 
an arbitrary distribution $h(t)$ of non-zero hopping matrix elements. For an RRG model with fixed hoppings $t=1$, as defined in Sec.~\ref{s2}, we have $h(t) = \delta(t-1)$. We will consider the generalized RRG ensemble (with random non-zero hoppings)  in the present subsection and will return to the ``standard'' version with $t \equiv 1$ starting from Sec.~\ref{s3.3}. In comparison with Refs.~\cite{mirlin1991universality,fyodorov1991localization}, where the SRM ensemble was considered, we have done the following modifications.
First, we have included the randomness in local potentials  characterized by the distribution $\gamma(\epsilon)$.  Second, we have imposed a condition of fixed connectivity, which is equal to $p$ not only in average (as in the SRM model) but also exactly for every vertex. 

Now we introduce the supersymmetric partition function, as in  Eqs. (\ref{superint}), (\ref{superaction}). Before averaging, the action reads
\be
\label{superaction-rrg}
\mathcal{L}_H(\Phi) =  - \frac{i}{2}\sum_{ij}\Phi_i^\dagger\hat{\Lambda}\left\{ \left [E+ \left (\frac{\omega}{2}+i\eta \right)\hat{\Lambda} \right]\delta_{ij}-H_{ij}\right\}\Phi_j,
\ee
where $\eta >0$ is an infinitesimal imaginary part of frequency. 
To get the partition function, we perform the averaging of the weight $e^{-\mathcal{L}_H(\Phi)}$  over the distribution  (\ref{rrg-distribution}). 
This generates the required ensemble of tight-binding Hamiltonians on RRG, with random diagonal and hopping matrix elements. 
In order to decouple the integrations over $A_{ij}$, we represent $\delta$-functions imposing the condition of fixed connectivity $p$ in Eq.~(\ref{rrg-distribution}) as 
 integrals over auxilary variables $x_i$, in analogy with the replica calculation in Ref.~\cite{PhysRevE.90.052109}.
The averaged partition function reads, in close analogy with Refs.~\cite{mirlin1991universality,fyodorov1991localization},
\bea
\langle Z \rangle & = & \int \prod_i d\Phi_i \frac{dx_i}{2\pi} e^{ipx_i} \exp\left\{ \sum_i \left[
\frac{i}{2}\Phi_i^\dagger\hat{\Lambda}(E - J_i\hat{K})\Phi_i \right. \right.
\nonumber \\
&+& \left.  \frac{i}{2}\left(\frac{\omega}{2} + i \eta \right)\Phi_i^\dagger\Phi_i  +  \ln \tilde{\gamma}(\frac{1}{2}\Phi_i^\dagger \hat{\Lambda} \Phi_i )\right]   
\nonumber \\
&+& \left.  \frac{p}{2N} \sum_{i \ne j}\left[ e^{-i(x_i+x_j)} \tilde{h} (\Phi_i^\dagger \hat{\Lambda} \Phi_j) - 1\right] \right\}.
\label{Z1}
\eea
Here the function $\tilde{\gamma}(z)$ is the Fourier transform of the distribution $\gamma(\epsilon)$ of energies, $\tilde{\gamma}(z) = \int d\epsilon\: e^{-i\epsilon z} \gamma(\epsilon)$. Further,  $\tilde{h}(z)$ is the Fourier transform of the distribution $h(t)$ of hoppings;  for an RRG model with fixed hoppings $t=1$ we have $\tilde{h}(z) = e^{-iz}$. Finally, $J_i$ in Eq.~(\ref{Z1}) are source fields; variation of $\langle Z \rangle$ with respect to them (and then setting them to zero) allows one to generate observables of interest. The corresponding matrix $\hat{K}$ is a diagonal matrix with elements equal to $+1$ for bosons and $-1$ for fermions. 

The next step is the decoupling of integrations over variables $\Phi_i$ associated with different sites. This is done by means of a functional generalization of the Hubbard-Stratonovich transformation:
\bea
&& \exp \left\{ \frac{p}{2N} \sum_{i \ne j}  e^{-i(x_i+x_j)} \tilde{h} (\Phi_i^\dagger \hat{\Lambda} \Phi_j) \right\} \nonumber \\
&& =  \int Dg \exp \left\{ - \frac{Np}{2} \int d\Psi d \Psi' g(\Psi) C(\Psi, \Psi') \right.
\nonumber \\ && \left. + p \sum_i e^{-ix_i} g(\Phi_i) \right\},
\eea
where $C(\Psi, \Psi')$ is a kernel of an integral operator inverse to that with the kernel $\tilde{h}(\Phi^\dagger \hat{\Lambda}\Psi)$. The integration $\int Dg$ runs over functions of a supervector $g(\Phi)$. Substituting this in Eq.~(\ref{Z1}) and performing integrations over $\Phi_i$, we obtain the expression for physical observables in terms of an integral over functions $g(\Phi)$:  
\be
\langle {\cal O} \rangle = \int Dg \: U_{\cal O}(g) e^{-N\mathcal{L}(g)}.
\label{g-funct-integral}
\ee
Here the action is (up to the factor $N$) 
\bea
\mathcal{L}(g) &=& \frac{m+1}{2}\int d\Psi d\Psi'g(\Psi)C(\Psi,\Psi')g(\Psi') \nonumber \\
&-& \ln\int d\Psi \: F^{(m+1)}_g(\Psi),
\label{Lg}
\eea
where we introduced $m = p-1$ and 
\bea
F^{(s)}_g(\Psi) &=& \exp  \left\{ \frac{i}{2} E\Psi^\dagger \hat{\Lambda} \Psi + 
\frac{i}{2}\left(\frac{\omega}{2} + i \eta \right)
\Psi^\dagger\Psi \right\} \nonumber \\
& \times & \tilde{\gamma}(\frac{1}{2}\Psi^\dagger \hat{\Lambda} \Psi ) g^s(\Psi).
\label{Fsg}
\eea
It is crucially important that the action in Eq.~(\ref{g-funct-integral}) is proportional to $N$. In the limit of large $N$, the integral in Eq. (\ref{g-funct-integral}) can be thus evaluated in the saddle-point approximation. The saddle-point configuration $g_0(\Psi)$ of the action  is determined by varying Eq.~(\ref{Lg}) with respect to $g$, which yields the equation
\be
\label{scvector}
g_0(\Psi)=\frac{\int d\Phi \: \tilde{h}(\Phi^\dagger \hat{\Lambda} \Psi) F^{(m)}_{g_0}(\Phi)}{\int d\Phi \: F^{(m+1)}_{g_0}(\Phi)}.
\ee
For symmetry reasons, the solution of Eq.~(\ref{scvector})  is a function of two invariants
\be
g_0(\Psi)=g_0(x,y); \quad x=\Psi^\dagger\Psi, \quad y=\Psi^\dagger \hat{\Lambda} \Psi.
\label{g0xy}
\ee
Because of the supersymmetry, the denominator in Eq.~(\ref{scvector}) is thus equal to unity, so that the saddle-point equation reduces to
\be
\label{sc1}
g_0(\Psi)= \int d\Phi \: \tilde{h}(\Phi^\dagger \hat{\Lambda} \Psi) F^{(m)}_{g_0}(\Phi).
\ee
We remind the reader that the function $\tilde{h}(z)$ becomes $h(z) = e^{-iz}$ for the model with non-random hoppings, as defined in Eq.~(\ref{H}). 

Equation (\ref{sc1}) is identical to the self-consistency equation describing the model on an infinite Bethe lattice, as derived within the supersymmetry formalism in Ref.~\cite{mirlin1991localization}. Equation (\ref{sc1}) can be reduced, by using
Eq.~(\ref{g0xy}), to a non-linear integral equation for $g_0(x,y)$.  It was further shown in Ref.~\cite{mirlin1991localization} that, for the case of a purely imaginary frequency ($\omega = 0$), the function $g_0(x,y)$ has an important physical interpretation. Specifically, it is  the Fourier-Laplace transform of the joint probability distribution $f^{(m)}(u', u'')$ of real and imaginary parts of  local Green function,
\be
\label{g0t}
g_0(x,y)=\int du'\int du'' f^{(m)}(u',u'')e^{\frac{i}{2}\left(u'y+iu''x\right)},
\ee
where $G_A^{(m)} (0, 0; E) = \langle 0 | (E- {\cal H} -i\eta)^{-1} | 0 \rangle = u' +i u''$ and a slightly modified lattice is considered, with the site $0$ having only $m$ neighbors. When rewritten in terms of $f^{(m)}(u',u'')$,  Eq.~(\ref{sc1}) becomes the self-consistency equation of Abou-Chacra et al., Ref.~\onlinecite{abou1973selfconsistent}.

A closely related object is the function $g_0^{(m+1)}(\Psi)$ which is expressed via $g_0(\Psi)$ as
\be
\label{g0m1}
g_0^{(m+1)}(\Psi)= \int d\Phi \: \tilde{h}(\Phi^\dagger \hat{\Lambda} \Psi) F^{(m+1)}_{g_0}(\Phi).
\ee
Again, for symmetry reason, it depends only on two invariants, $g_0^{(m+1)}(\Psi) = g_0^{(m+1)}(x,y)$. The function $g_0^{(m+1)}(x,y)$ is  the Fourier-Laplace transform of the joint probability distribution $f^{(m+1)}(u', u'')$ of real and imaginary parts of  local Green function at any site of the undeformed Bethe lattice. 

The applicability of the saddle-point method at $N \gg 1$, which allows one to reduce the evaluation of correlation functions to the solution of the self-consistency equation and calculation of integrals involving the saddle-point function, is a key to the analytical progress for the RRG model as shown below. We will use analytical results concerning the self-consistent solution already known from previous studies of models on an infinite Bethe lattice and some of their generalizations. Furthermore, the self-consistency equation can be efficiently solved numerically via the pool method (population dynamics), which complements the purely analytical approach; we will employ this below for various observables of interest.

 Evaluating the integral by the saddle-point method, it is crucially important to take into account that the Anderson localization transition is a spontaneous symmetry breaking phenomenon with $g_0(x,y)$ being the functional order parameter. The saddle-point equation is invariant, at $\omega=0$ and $\eta\to +0$, under transformations $g(\Psi) \to g(\hat{T}\Psi)$ with $\hat{\overline{T}} \hat{\Lambda}\hat{T} = \hat{\Lambda}$, where $\hat{\overline{T}}$ is the matrix that rotates the conjugate vector $\Psi^\dagger$. The proper choice of the manifold of matrices $\hat{T}$ is a rather non-trivial question that has been extensively discussed in the literature in the context of derivation of the supersymmetric $\sigma$ model; see, in particular, Refs.~ \cite{verbaarschot1985grassmann,efetov1999supersymmetry,mirlin00}. In brief, the manifold is defined by $\hat{T}^\dagger \hat{L} \hat{T} = \hat{L}$, where $\hat{T}^\dagger$ is the hermitian conjugate of $\hat{T}$, and $\hat{L}$ is the diagonal matrix equal to unit matrix in the fermionic subspace and to $\hat{K}$ in the bosonic subspace. 
The matrix $\hat{\overline{T}}$ introduced above is related to $\hat{T}^\dagger$ via $\hat{\overline{T}} = \hat{\Lambda} \hat{L} \hat{T}^\dagger \hat{\Lambda} \hat{L}$. 

In the localized phase, the symmetry of the saddle point solution is that of the equation, $g_0(x,y)=g_0(y)$. More accurately, dependence of the solution $g_0(x,y)$ of the self-consistency equation  on the variable $x$ appears only on the scale $\sim \eta^{-1}$ due to the term with $\eta$ in the action that explicitly breaks the symmetry.  Thus, in the localized phase, the integral (\ref{g-funct-integral}) is determined by a single saddle point $g_0$. 
This remains true at the critical point as well. On the other hand, in the delocalized phase, dependence on the variable $x$ persists, in the limit of $\omega, \eta\to 0$, signifying spontaneous symmetry breaking. As a result, a manifold of saddle-points emerges, parametrized by matrices $\hat{T}$:
\be
g_{0T}(\Psi) = g_0(\Psi^\dagger\hat{\overline{T}} \hat{T}\Psi, \: \Psi^\dagger\hat \Lambda \Psi). 
\label{g0T}
\ee
The  integral (\ref{g-funct-integral}) in this situation becomes an integral over the manifold of saddle points, or, equivalently, the manifold of $\hat{T}$-matrices. Specifically, it is important to integrate over the whole manifold  when the frequency $\omega$ (which breaks the symmetry with respect to rotations by matrices  $\hat{T}$ generating the manifold) is small. This is, in particular, the case for auto-correlation of an eigenfunction ($\omega=0$), as well for correlations of sufficiently close in energy eigenfunctions and energy levels. On the other hand, at sufficiently high energies integrals are dominated by the single saddle point $g_0$. 

In the delocalized phase, the dependence of the function $g_0(x,y)$ on the variable $x$ appears on a scale $x \sim N_\xi$ which diverges exponentially when the system approaches the transition point (critical disorder) $W_c$:
\be
\ln N_\xi \sim (W_c - W)^{-1/2}.
\label{Nxi}
\ee
The notation $N_\xi$ emphasizes that this scale has a meaning of the correlation volume
associated with the correlation length $\xi$:
\be 
N_\xi \sim m^\xi \,.
\label{Nxi1}
\ee  

Let us now turn to the case of $n$-orbital model with $n\gg 1$ which is described by the supermatrix $\sigma$-model action (\ref{action}). In full analogy with the above discussion of the supervector theory for $n=1$, one can perform the structural averaging over the RRG ensemble, which reduces the theory (\ref{action}) to an integral over functions $F(Q)$. The corresponding  self-consistency equation has the form
\be
\label{sc}
g_0(Q)=\int e^{-\str\left[-2J(Q-Q')^2+\frac{\pi\eta}{\delta_0}\Lambda Q'\right]}g_0^{m}(Q')DQ'
\ee
and is identical to the self-consistency equation for the $\sigma$ model on an infinite Bethe lattice\cite{efetov1985anderson,zirnbauer1986localization,zirnbauer1986anderson,efetov1987density,efetov1987anderson,verbaarschot1988graded}.
In the localized phase (and at the critical point), the solution (in the limit $\eta \to 0$)  is  $g_0(Q)=1$, which corresponds to preserved symmetry. In the delocalized phase, the symmetry is broken, and the solution is a non-trivial function of $Q$. As a result a manifold of saddle-points arises parametrized by matrices $\hat{T}$, in full analogy with the $n=1$ model:  $g_T(Q) = g_0 (\hat{T}^{-1}Q\hat{T})$.  In analogy with the supervector ($n=1$) formulation, the function $g_0(Q)$ depends, for symmetry reason, only on two scalar invariants (eigenvalues of the boson-boson block of $Q$), $\lambda_1$ and $\lambda_2$. Of central importance is dependence on $\lambda_1$, which appears on the scale $\sim \eta^{-1}$ in the localized phase and on the emerging scale $N_\xi$ in the delocalized phase---again, in perfect correspondence to the $n=1$ model. Furthermore, the correlation volume $N_\xi$ shows the same exponential critical behavior (\ref{Nxi}). 

It is known from previous works on models in finite-$d$ and on the Bethe lattice that various observables  characterizing the disordered system have very similar behavior in the cases of the $n=1$ model and the $\sigma$ model (corresponding to $n\gg 1$). We thus expect such a close similarity between these models also for eigenfunction and energy level statistics on RRG. Indeed, our results presented below support this expectation. We have identified, however, a difference between two models in subleading (logarithmic) factors in $\omega$-dependence of $\beta_E(0,\omega)$  at criticality, see Sec.~\ref{sec:different_wavefunctions} below.

\subsection{Single wavefunction}
\label{s3.3}

From now on we focus on a RRG model with all non-zero hopping matrix elements fixed to $t=1$. 

In this subsection, we consider self-correlations of eigenfunctions, as encoded in the correlation function between $\alpha_E(i,j)$ defined by Eq.~(\ref{alphadef}). For brevity, we will drop the subscript $E$ and use the distance between the sites $i$ and $j$ as the argument, thus denoting this correlation function as $\alpha(r)$. 
The correlation function $\alpha(r)$  can be expressed in terms of Green functions as follows
\be
\label{def}
\alpha(r) = \frac{1}{\pi\nu N} \lim_{\eta\to 0}\eta\left<G_R(i,i)G_A(j,j)\right>,
\ee
where $G_{R,A} (j,j) = \langle j | (E- {\cal H} \pm i\eta)^{-1} | j \rangle$.
The product of Green functions entering Eq.~(\ref{def}) can, in turn, be written, for a given realization of the Hamiltonian, as a superintegral,
\bea
\label{Gsuper}
G_R(i,i)G_A(j,j) &=& \frac{1}{16}\int\prod_{k}[d\Psi_k] \left(\Psi_1^+\hat K\Psi_1\right)\left(\Psi_2^+\hat K\Psi_2\right) \nonumber \\
& \times &  e^{-\mathcal{L}_H(\Psi)},
\eea
with action defined by Eq.~(\ref{superaction-rrg}). 

We consider first the single-point correlation function $\alpha(0)$ (i.e., $i=j$); this is the same (up to a factor $N$) that the average IPR: $P_2 = N\alpha(0)$.   The calculation of $\alpha(0)$ for the SRM ensemble was carried out in Ref.~\cite{fyodorov1991localization} (for technical details see Ref.~\cite{fyodorov1992novel}); we follow the same route for the RRG model (which requires only minor modifications). Performing the ensemble averaging as outlined in Sec.~\ref{s3.2}, we obtain the averaged product of the Green functions at coinciding points in the form of the functional integral (\ref{g-funct-integral}),
\be
\label{gsuper}
\left<G_R(j,j) G_A(j,j)\right>= \int Dg \: U(g) e^{-N\mathcal{L}(g)},
\ee
with the action $\mathcal{L}(g)$ given by Eq.~(\ref{Lg}) and
\be
\label{usuper}
U(g)=\int  [d\Psi] \: \frac{1}{16} \left(\Psi_1^\dagger\hat K\Psi_1\right)\left(\Psi_2^\dagger\hat K\Psi_2\right)  F_g^{(m+1)}(\Psi).
\ee
In the limit of large $N$, the integral in Eq. (\ref{gsuper}) is evaluated by the saddle-point method as explained in Sec.~\ref{s3.2}. 

In the localized regime, $W >  W_c$, the integral (\ref{gsuper}) is dominated by a single saddle point $g_0(x,y)$.  Integration over fluctuations around the saddle point (``massive modes'') gives unity due to supersymmetry, so that Eq. (\ref{gsuper}) reduces to $U(g_0)$. The calculation of $U(g_0)$ as given by the integral (\ref{usuper}) yields 
$U(g_0)=C/\eta$, with $C$ a constant of order unity. 
The factor $\eta^{-1}$ (divergent in the limit $\eta\to 0$) arises from the integration in Eq. (\ref{usuper}): the convergence of this integration is provided by the fast decay of the saddle-point configuration $g_0(x,y)$ at $x\gtrsim\eta^{-1}$, as set by a symmetry-breaking term proportional to $\eta$. 
According to Eq.~(\ref{def}), we thus find 
\be
\alpha(0) = C / \pi \nu N; \qquad C \sim 1,
\label{alpha0loc} 
\ee
which corresponds to the expected behavior of $P_2$ in the localized phase, cf. Eq.~(\ref{P2sc}).
This result is identical to the one found on an infinite Bethe lattice\cite{mirlin1991localization}, which is not surprising: localized states are insensitive to the boundary conditions. The expression of the constant $C$ in terms of the function $g_0(x,y)$ can be found Ref. \onlinecite{mirlin1991localization}.   It is important that $C$ has a non-zero limit at the critical point, $W = W_c$: from the IPR point of view, critical states on RRG are very similar to localized ones, as was already discussed in Sec.~\ref{s3.1}.
 
In the delocalized phase, $W < W_c$, the manifold of the saddle points $g_{0T}(\Psi)$, Eq.~(\ref{g0T}), contributes and the integral in Eq.~(\ref{gsuper}) 
takes the following form:
\be
\int Dg e^{-N\mathcal{L}(g)}U(g)=\int d\mu(\hat T) \: U(g_{0T}) \: e^{-\frac{\pi}{2}N\eta\nu\mathrm{Str}\left[\hat{\overline{T}}\hat T\right]}.
\label{IPR-T-integral}
\ee
The supertrace in the exponent of Eq.~(\ref{IPR-T-integral}) can be written in the form  $\mathrm{Str}\: \hat{\overline{T}}\hat T = \mathrm{Str}\: Q \hat \Lambda$, where $Q = \hat T^{-1} \hat \Lambda \hat T$ is the $\sigma$-model field ($Q^2 = 1$) corresponding to the zero mode $\hat T$. 
The factor $U(g_{0T})$ in Eq.~(\ref{IPR-T-integral}) is given by the integral (\ref{usuper}). A convenient way to evaluate this integral is to use the analog of Eq.~(\ref{g0m1}) for $g_{0T}^{(m+1)}$ and expand it up to terms of fourth order in $\Phi$. This yields, in full analogy with Ref.~\cite{fyodorov1992novel}, 
\bea
U(g_{0T}) &=&  -\frac{1}{64} \left[g_{0,xx}^{(m+1)} \! \left( \mathrm{Str}\: Q \hat K \frac{1+\hat \Lambda}{2}\ \mathrm{Str}\: Q \hat K \frac{1-\hat \Lambda}{2} \right. \right.
\nonumber \\
&+& 2 \left. \left. \mathrm{Str}\: Q \hat K \frac{1+\hat \Lambda}{2} Q \hat K \frac{1-\hat \Lambda}{2} \right)  - g_{0,yy}^{(m+1)}
\right],
\label{Ug0T}
\eea
where $g_{0,xx}^{(m+1)} = \partial^2_x g_0^{(m+1)}|_{x,y=0}$ and $g_{0,yy}^{(m+1)} = \partial^2_y g_0^{(m+1)}|_{x,y=0}$. Performing the zero-mode integration (\ref{IPR-T-integral}), one finds that the term proportional to $g_{0,xx}^{(m+1)}$ in Eq.~(\ref{Ug0T}) yields a contribution that diverges as $\eta^{-1}$ at $\eta \to 0$.  After substitution in Eq.(\ref{def}), this yields 
\be
\label{alpha0}
\alpha(0)=\frac{12}{N^2}\frac{g_{0,xx}^{(m+1)} }{\pi ^2\nu^2}.
\ee

The coefficient $g_{0,xx}^{(m+1)}$ in Eq.~(\ref{alpha0}) has an important physical meaning. Since the function $g_0^{(m+1)}(x,y)$ is the Fourier-Laplace transform of the distribution of local Green functions on an infinite Bethe lattice (see Sec.~\ref{s3.2}), $g_{0,xx}^{(m+1)}$ is proportional to the average square of the local density of states $\nu(j) = - (1/\pi) {\rm Im}\: G(j,j)$:  
\be
g_{0,xx}^{(m+1)} = (\pi^2 / 4) \langle \nu^2 \rangle_{\rm BL} \:.
\label{nu2}
\ee
The subscript ``BL'' here indicates that the average should be computed with the help of solution of the self-consistency equation that describes  the model on an infinite Bethe lattice. Equation (\ref{alpha0}) can thus be written in the form
\be
\label{alpha0-1}
\alpha(0)=\frac{3}{N^2}\frac{\left<\nu^2\right>_{\rm BL}}{\nu^2}.
\ee

Behavior of the coefficient (\ref{nu2}), which is determined by properties of the self-consistent solution,  is well understood\cite{mirlin1991localization,fyodorov1991localization,fyodorov1992novel}.
Deeply in the delocalized phase ($W \ll W_c$), the fluctuations of the local density of states are weak, $\langle \nu^2 \rangle_{\rm BL} / \nu^2 \simeq 1$, so that Eq.~(\ref{alpha0-1}) reduces (up to small corrections) to the result of the Gaussian orthogonal ensemble. 
On the other hand, when $W$ approaches the critical value $W_c$, these fluctuations become strong,  $\langle \nu^2 \rangle_{\rm BL} / \nu^2 = N_\xi$, where 
$N_\xi$ is the correlation volume that shows the critical behavior (\ref{Nxi}). In principle, $N_\xi$ is defined up to a coefficient of order unity; we can use the last formula to fix this coefficient. Thus, in the delocalized phase
\be
\label{alpha02}
\alpha(0)=3N_\xi/N^{2}.
\ee
Comparing this with the behavior in the critical point, Eq.~(\ref{alpha0loc}), we see that they match when the number of sites is of the order of correlation volume, $N \sim N_\xi$, as expected. For $N \gg N_\xi$ the system is in the delocalized regime, and Eq.~(\ref{alpha02}) holds. On the other hand, for $N \ll N_\xi$ the system is effectively at criticality, and $\alpha(0)$ is given by Eq.~(\ref{alpha0loc}).  The $1/N$ scaling of the  IPR $P_2 = N\alpha(0)$ in the delocalized phase at $N \gg N_\xi$ as given by Eq.~(\ref{alpha02}) is a manifestation of ergodicity of delocalized states in the RRG model. This scaling was numerically demonstrated in 
Ref.~\onlinecite{tikhonov2016anderson} for the RRG model and in  Ref.~\cite{garcia-mata17} for a model with fluctuating coordination number that is intermediate between RRG and SRM.

Let us now turn to the behavior of  $\alpha(r)$ as a function of distance $r$. For this purpose, we should perform averaging in Eq. (\ref{alphadef}) over all pairs of RRG nodes such that the shortest path between them has the length $r$. When doing so, we should fix $r$ and consider sufficiently large $N$ such that the distance between a random pair of sites is almost surely larger than $r$. The corresponding condition is $r < \kappa \log_m N$, with $\kappa < 1$. Under this condition, the requirement that the shortest path between two (otherwise random) sites has length $r$ is equivalent to the requirement that there exists a path of length $r$ connecting these two sites. To impose the latter  requirement, we modify the probability distribution (\ref{rrg-distribution}) defining the RRG ensemble in the following way. We fix $r+1$ sites (that we label $0,1, \ldots, r$) and, for each of the pairs $(i,j)$ from the set $\{(0,1), (1,2), \ldots, (r-1,r)\}$ we replace the factor  $( 1 - p/N) \delta(A_{ij}) + (p/N) \delta(A_{ij} - 1)$ in Eq.~(\ref{rrg-distribution}) by $\delta(A_{ij}-1)$. This implies that there is a path of length $r$ (which goes through the sites $1,2, \ldots, r-1$) connecting the sites $0$ and $r$.  After this, it remains to evaluate the correlation function (\ref{def}), choosing the site $0$ as $i$ and the site $r$ as $j$. 

The further calculation proceeds along the same way as for $\alpha(0)$. 
The averaged product of Green functions $\left<G_R(0,0) G_A(r,r)\right>$  is obtained in the same form (\ref{gsuper}) of the integral over functions $g(\Psi)$, where now 
\bea
\label{usuper-r}
U(g) &=& \int  \prod_{i=0}^r [d\Psi_i] \: \frac{1}{16} \left(\Psi_{0,1}^\dagger\hat K\Psi_{0,1}\right)\left(\Psi_{r,2}^\dagger\hat K\Psi_{r,2}\right)  \nonumber \\
& \times &
F_g^{(m)}(\Psi_0) \prod_{j=0}^{r-1} e^{-i\Psi_j^\dagger \Psi_{j+1}} \prod_{k=1}^{r-1} F_g^{(m-1)}(\Psi_k)   \nonumber \\
& \times & F_g^{(m)}(\Psi_r).
\eea
The integral is again evaluated by means of the saddle-point method. Since the action to the order $N$ is the same as before, the saddle-point equation for the function $g_0$ remains unchanged. 

\begin{figure*}[tbp]
\minipage{0.5\textwidth}\includegraphics[width=\textwidth]{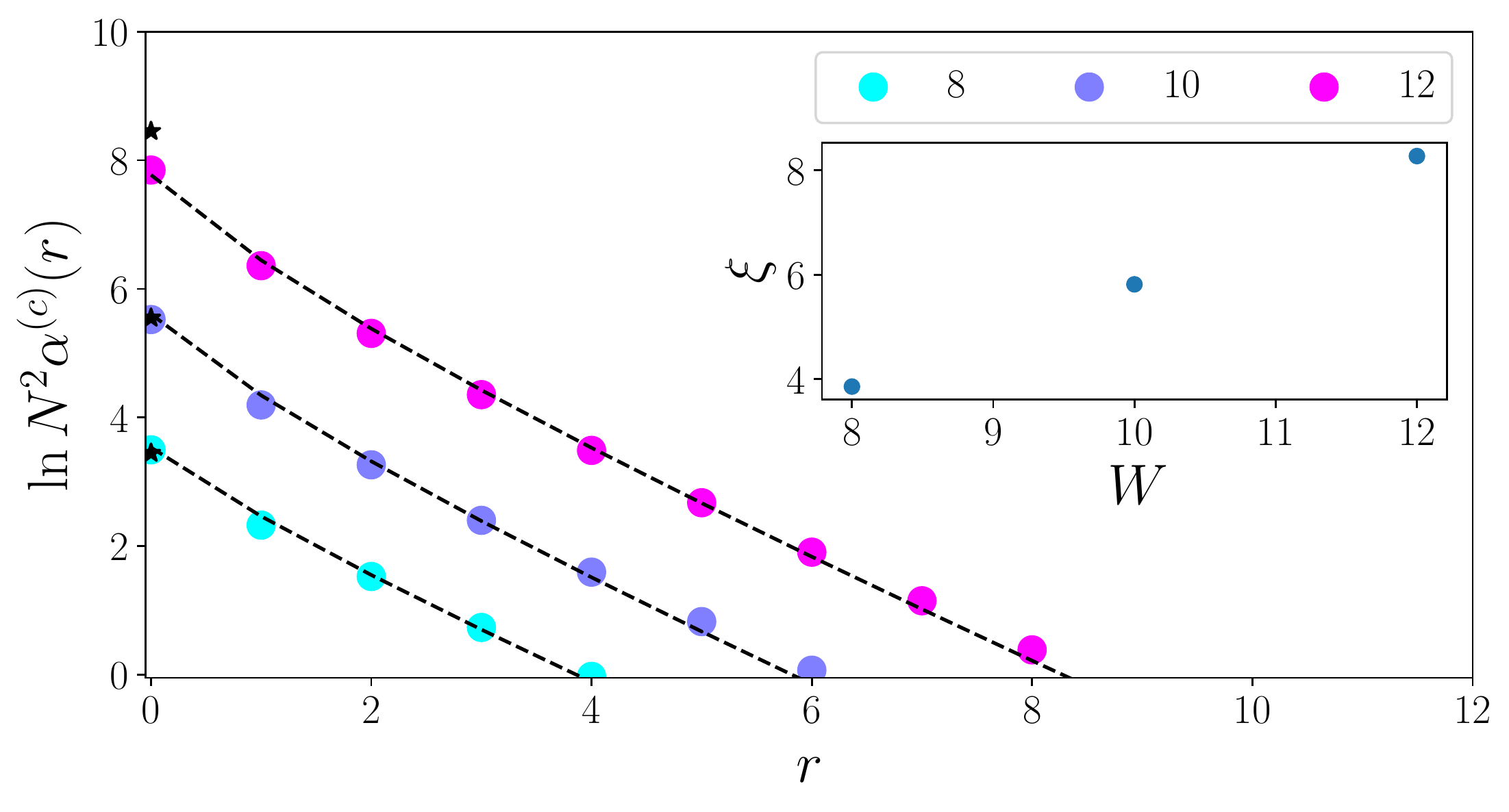}\endminipage
\minipage{0.5\textwidth}\includegraphics[width=\textwidth]{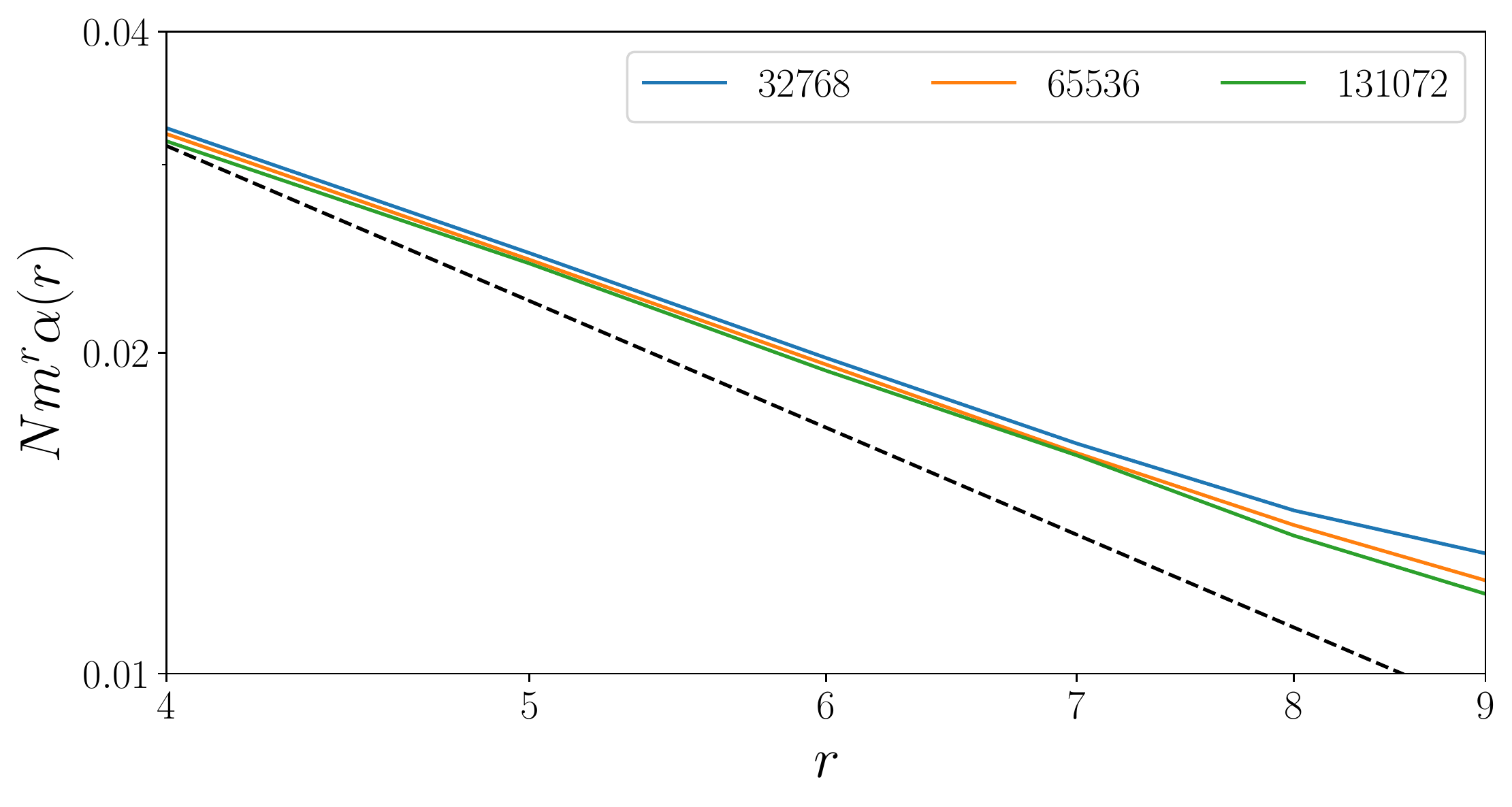}\endminipage
\caption{Eigenfunction self-correlations $\alpha(r)$. Left: connected part $N^2\alpha^{(c)}(r) = N^2\alpha(r) -1$ in the delocalized phase for $N=131072$ and disorder values $W=8$, 10, and 12. Dashed black lines: fit to $\ln N^2\alpha^{(c)}(r)=-r\ln m-c_1^{(\alpha)}\ln (r+1)-c_2^{(\alpha)}$, see Eq.~(\ref{alphadeloc}). For values of $c_1^{(\alpha)}$, see text. 
Inset: correlation length $\xi(W)$ as estimated from the condition $\ln N^2\alpha^{(c)}(r)=0$. Star symbols: values of $\alpha(0)$ as obtained by evaluating Eq.(\ref{alpha0-1}) by means of population dynamics. Inset: $\xi(W)$ as obtained from the condition $\ln N^2\alpha^{(c)}(r)=0$. 
Right:  $N m^r\alpha(r)$ at the critical point $W=18$.  Black dashed line: $c / r^{3/2}$, see Eq.~(\ref{alphadeloc}).}
\label{same}
\end{figure*}

In the localized phase, we have a single saddle point $g_0$ which is the solution of the Bethe-lattice self-consistency equation.  The integral (\ref{usuper-r}) is then nothing but the expression for the correlation function $\langle \frac{1}{16} (\Psi_{0,1}^\dagger\hat K\Psi_{0,1})(\Psi_{r,2}^\dagger\hat K\Psi_{r,2})\rangle_{\rm BL}$ on an infinite Bethe lattice.  We recall that, to the order that we consider (averaged products of two Green functions), there are two different non-trivial correlation functions on an infinite Bethe lattice:
\bea
\label{K1}
K_1(r) &=& \langle G_R(i,i) G_A(j,j) \rangle_{\rm BL}  \nonumber \\ & = & 
  \langle \frac{1}{16} (\Psi_{i,1}^\dagger\hat K\Psi_{i,1})(\Psi_{j,2}^\dagger\hat K\Psi_{j,2})\rangle_{\rm BL}; \\
K_2(r) &=& \langle G_R(i,j) G_A(j,i) \rangle_{\rm BL} \nonumber \\ & = & 
 \langle \frac{1}{16} (\Psi_{j,1}^\dagger\hat K\Psi_{i,1})(\Psi_{i,2}^\dagger\hat K\Psi_{j,2})\rangle_{\rm BL}.
 \label{K2}
\eea
Here $r$ is the distance between the sites $i$ and $j$. The correlation functions of the type $\langle G_R G_R \rangle$ and $\langle G_A G_A \rangle$ are not sensitive to the Anderson transition and decouple into products of averaged Green functions. For the $\sigma$ model on an infinite Bethe lattice, direct counterparts of these correlation functions are
\bea
\label{sigmaK1}
K_1(r)=\left<Q_{bb}^{11}\left(i\right) Q_{bb}^{22}(j)\right>_{\rm BL};\\
K_2(r)=\left<Q_{bb}^{12}\left(i\right) Q_{bb}^{21}(j)\right>_{\rm BL}.
\label{sigmaK2}
\eea
Here the upper indices of (unitary-class) $Q$ matrices refer to retarded-advanced decomposition and the subscripts ``bb" to the boson-boson elements. 
The $r$ dependence of the Bethe-lattice correlation functions $K_1(r)$ and $K_2(r)$ has been studied in detail, both for the $\sigma$ model 
\cite{zirnbauer1986localization,zirnbauer1986anderson,efetov1987density,efetov1987anderson,verbaarschot1988graded} 
(also in the ``toy version''\cite{gruzberg96}) and the $n=1$ model\cite{mirlin1991localization}.  Essentially the same behavior was found for the $n=1$ model and for large-$n$ models ($\sigma$ models) of all three Wigner-Dyson classes.
Specifically, the results (for $\omega = 0$ and $\eta \to 0$) are as follows. In the localized phase, $W > W_c$, the correlation functions $K_1(r)$ and $K_2(r)$ have $1/\eta$ singularity, are equal to each other, and decay with $r$ as 
\be
\label{K-localized}
K_1(r) = K_2(r) \sim \frac{1}{\eta} m^{-r} e^{-r/\zeta} r^{-3/2},
\ee
where $\zeta$ is the localization length that diverges at the critical point as $\zeta \sim (W-W_c)^{-1}$. In the delocalized phase, $W < W_c$, close to the transition point (where the correlation volume $N_\xi$ is large), the result for the function $K_2(r)$ reads
\be
\label{K2-deloc}
K_2(r) \sim N_\xi m^{-r}  r^{-3/2}.
\ee
The function $K_1(r)$ in this regime shows the same behavior as $K_2(r)$ for not too large distances $r < \xi$, whereas for $r > \xi$ it saturates at the value given by its disconnected part $K_{1}^{(d)}$ equal to $K_{1}^{(d)} = |\langle G_R(j,j) \rangle |^2$ for Eq.~(\ref{K1}) and $K_{1}^{(d)} = 1$ for Eq.~(\ref{sigmaK1}).  In other words\cite{foot1}
\be
\label{K1-deloc}
K_1(r) \simeq K_2(r) + K_{1}^{(d)}.
\ee

We use now these Bethe-lattice correlation function for our analysis of eigenfunction correlations on RRG. 
In the localized phase,  $W > W_c$, $U(g)$ is simply $\langle \frac{1}{16} (\Psi_{0,1}^\dagger\hat K\Psi_{0,1})(\Psi_{r,2}^\dagger\hat K\Psi_{r,2})\rangle_{\rm BL}$, which is nothing but the correlation function $K_1(r)$, so that
\be
\label{P2K1loc}
\alpha(r)=\frac{1}{\pi\nu N}\lim_{\eta\to 0}\eta K_1(r,\eta).
\ee
 Using Eq.~(\ref{K-localized}), we immediately find
\be
\label{alpha-r-loc}
\alpha(r) \sim \frac{1}{N}m^{-r} e^{-r/\zeta} r^{-3/2}.
\ee
This result is extended to the critical point by setting $\zeta = \infty$, which yields
\be
\label{alphaloc}
\alpha(r)\sim \frac{1}{N}\frac{m^{-r}}{r^{3/2}}.
\ee

In the delocalized phase, we have a manifold of saddle-points $g_{0T}$ and should evaluate the integral over the zero mode $\hat T$, see Eq.~(\ref{IPR-T-integral}). Evaluating the integral (\ref{usuper-r}) on a saddle point $g = g_{0T}$ we find the same $Q$-dependent structures as those that emerged in Eq. (\ref{Ug0T}), see two terms in the bracket multiplying $g_{0,xx}^{(m+1)}$. The first of them has now as a prefactor the Bethe-lattice correlation function $K_1(r)$ (plus disconnected terms of the $\langle G_R G_R \rangle$ and $\langle G_A G_A \rangle$ type), while the second one is multiplied by the correlation function $K_2(r)$. Performing the zero-mode integration (\ref{IPR-T-integral}) and substituting the result into Eq.~(\ref{def}), we get 
\bea
\label{alpha-r-deloc}
\alpha(r) &=& \frac{1}{2\pi^2 N^2} [K_1 (r) + 2 K_2(r)  \nonumber \\ 
&-& {\rm Re} \langle G_R(0) \rangle^2 - 2{\rm Re} \langle G_R(r) \rangle^2 ].
\eea
The last term in square brackets in Eq.~(\ref{alpha-r-deloc}) is always small in comparison with the sum of other terms in the regime of large $N_\xi$ that we are focusing on.  Using Eqs.~(\ref{K1-deloc}) and (\ref{K2-deloc}), we find the behavior of the eigenfunction self-correlation in the delocalized phase of the RRG model:
\be
\label{alphadeloc}
\alpha(r)\sim \frac{N_{\xi}}{N^2}\frac{m^{-r}}{r^{3/2}}, \qquad r < \xi.
\ee
For $r > \xi$ the correlation function $\alpha(r)$ is governed by disconnected  parts in Eq.~(\ref{alpha-r-deloc}),  yielding $\alpha(r) \simeq 1$.
\cite{foot2} 
 The analysis can be extended to the $\sigma$ model (i.e., large-$n$ model) on RRG. The results are fully analogous to those of the $n=1$ model. Indeed, as shown above, the results are expressed in terms of the infinite-Bethe-lattice correlation functions $K_1(r)$ and $K_2(r)$, which have qualitatively the same behavior in the $n=1$ model and the $\sigma$ model. 

\begin{figure}[tbp]
\includegraphics[width=0.5\textwidth]{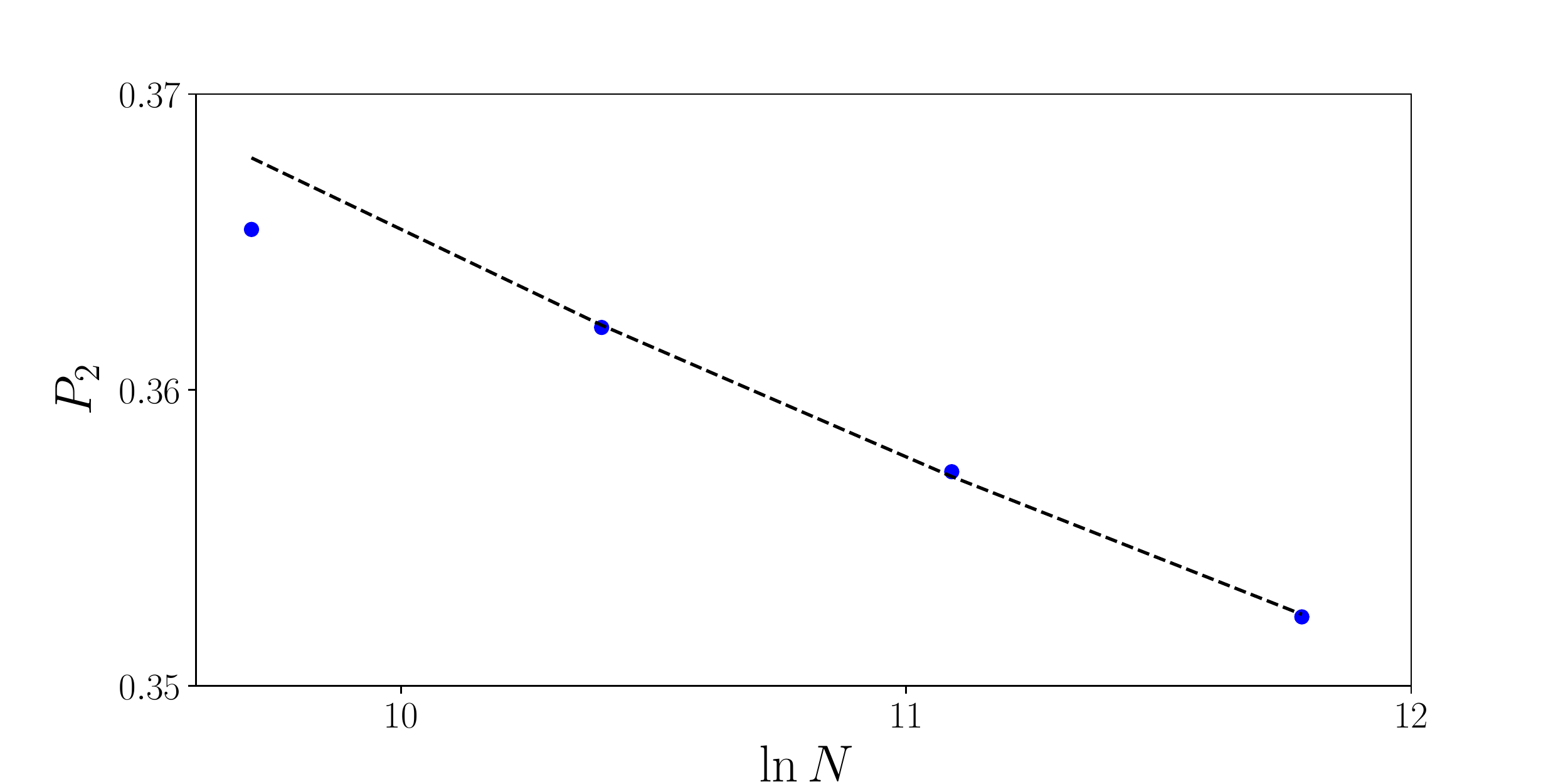}
\caption{Inverse participation ratio at the critical point ($W=18$) as a function of the system size as found by exact diagonalization for system sizes $N=16384$, $32768$, $65536$, and $131072$. Dashed line: fit with $P_2=c_1^{(P)}+c_2^{(P)}/\ln^{1/2}N$, where $c_1^{(P)}\simeq 0.20$.}
\label{p2loc}
\end{figure}

We can now compare the results of exact analysis,  Eqs. (\ref{alphaloc}) and (\ref{alphadeloc}), with the conjecture (\ref{P2sc2}) obtained by means of an extrapolation of finite-$d$ results to the $d \to \infty$ limit. We see that Eq.~(\ref{P2sc2}) correctly captures the leading behavior of the correlation function $\alpha(r)$. On the other hand, the accurate treatment  produces also a subleading factor $r^{-3/2}$ in Eqs. (\ref{alphaloc}) and (\ref{alphadeloc}).

Now we confront the analytical predictions for $\alpha(r)$ with results of the exact diagonalization of the RRG ensemble.
In Fig.~\ref{same}  (left panel) we show the connected part $\alpha^{(c)}(r) = \alpha(r) - 1/N^2$ evaluated numerically (by exact diagonalization) for  RRG with coordination number $m+1 = 3$, matrix size $N= 131072$  and several values of disorder, $W = 8, \: 10, \: 12$ corresponding to the delocalized phase. The position of the Anderson transition for this model is known to be $W_c \simeq 17.5$, so that these value of $W$ are sufficiently close to $W_c$ in the sense that the correlation volume is large, $N_\xi \gg 1$. At the same time, they are not too close, so that the condition $N_\xi \ll N$ is also met. 
The observed dependence of $\ln(N^2\alpha_c(r))$ on $r$  is nearly linear, in agreement with the analytically predicted exponential decay, Eq.~(\ref{alphadeloc}).
To check the subleading power-law factor $r^{-3/2}$, we have fitted the data to $\ln N^2\alpha_c(r)=-r\ln m-c_1^{(\alpha)}\ln (r+1)-c_2^{(\alpha)}$, with $c_1^{(\alpha)}$ as a free parameter. 
The fits are shown in the figure by dashed lines. The extracted values of $c_1^{(\alpha)}$ are $0.54,\;0.82,\;0.92$ for $W= 8$, 10, and 12, respectively. We see that $c_1^{(\alpha)}$ is smaller than $3/2$ but increases with increasing correlation length $\xi$, in consistency with the analytic prediction of the power-law exponent 3/2 in the asymptotic limit of large $\xi$.  The values of the correlation length $\xi$ as extracted from the condition $\ln(N^2\alpha^{(c)}(r)) = 0$ are shown in the inset of Fig.~\ref{same}. They nicely match the results for $\xi(W)$ as estimated in 
Ref.~\onlinecite{tikhonov2016anderson} from the analysis of the IPR scaling with $N$, where $\xi$ was found to increase from $\xi \simeq 9$ to $\xi \simeq 16$ in the range of disorder $W$ from 13 to 16. 

To further verify  the analytical prediction of the dependence of $\alpha(r)$ at criticality ($W=W_c$), we have performed exact-diagonalization analysis also  for $W=18$. (This value is so close to $W_c$ that the system is critical at any system size $N$ for which the numerical analysis is feasible.) First, we consider the behavior of the IPR $P_2=N\alpha(0)$, for which Eq. (\ref{alpha0loc}) predicts a limiting value of order unity at $N\to\infty$.  In Fig. \ref{p2loc}  the dependence of the IPR on the system size is shown. In agreement with analytical prediction, the IPR is only weakly dependent on the system size $N$. 
We fit the size dependence by the following formula:
\be
\label{P2scalingloc}
P_2=c_1^{(P)}+ \frac{c_2^{(P)}}{\ln^{1/2}N}
\ee
to extrapolate to the limit $N\to\infty$. This form of the $N$-dependence will be motivated in Sec. \ref{subsubsec:samepoint} below. 
This yields the limiting $N\to\infty$ value $P_2 \simeq 0.20$. 
In Sec. \ref{subsubsec:samepoint} we will also present an alternative way (based on the population dynamics) to calculate the limiting value of the critical IPR and will demonstrate a good agreement between the two approaches. 

The result for $\alpha(r)$ at finite $r$ is also in a very good agreement with the analytical prediction (\ref{alphaloc}). To demonstrate this, we show in the right panel of Fig.~\ref{same}  the product $Nm^r\alpha(r)$ as a function of $r$ on the double-logarithmic scale. We see that lines almost collapse, in agreement with $1/N$ dependence of $\alpha(r)$ at criticality, Eq.~(\ref{alphaloc}). A small drift is related to finite-size corrections. With increasing length $N$, the curves approach the straight line with the slope 3/2, as predicted. 

It should be emphasized  that Eqs.~(\ref{alpha0-1}) and  (\ref{alpha-r-deloc}) yield exact relations (i.e., without any unknown factors of order unity) between the eigenstate correlation functions on RRG (in the delocalized phase with $N \gg N_\xi$) and correlation functions of the infinite-Bethe lattice model. The latter correlation functions can be obtained numerically. In this discussion, we restrict ourselves to the case $r=0$ when the correlation function of interest [the one in the right-hand-side of Eq.~(\ref{alpha0-1})] is obtained directly from the solution of the self-consistency equation for $g_0(x, y)$, Eq. (\ref{scvector}). (For $r >0$ one further needs to evaluate an integral involving an $r$-th power of a certain integral operator.)  An efficient  approach to solution of such self-consistency equations is known as pool method (or, equivalently, population dynamics), see, e.g. Ref.~\cite{abou1973selfconsistent,biroli2010anderson}. This method amounts to solving the corresponding equation in its distributional form, which is obtained under transformation according to Eq.~(\ref{g0t}). In terms of the distribution function $f^{(m)}$ defined in 
Eq.~(\ref{g0t}) the self-consistency equation reads 
\bea
f^{(m)}(u) &=& \int d\epsilon\: \gamma(\epsilon)  \left (\prod_{r=1}^{m} du_r \, f^{(m)}(u_r)  \right) \nonumber \\
 & \times & \delta\left ( u - \frac{1}{E+i\eta -\epsilon - \sum_{r=1}^{m} {u_r}} \right).  
\label{fm}
 \eea
Further, the distribution function $f^{(m+1)}$ defined below Eq.~(\ref{g0m1}) is expressed through $f^{(m)}$ as 
\bea
f^{(m+1)}(u) &=& \int d\epsilon\: \gamma(\epsilon)  \left (\prod_{r=1}^{m+1} du_r \, f^{(m)}(u_r)  \right) \nonumber \\
 & \times & \delta\left ( u - \frac{1}{E+i\eta -\epsilon - \sum_{r=1}^{m+1} {u_r}} \right).  
\label{fm1}
 \eea
Here $u$ is a complex variable representing the local retarded Green function, $G(j,j; E+i\eta) = u = u^{'} - iu^{''}$.
Equivalently, these equations can be presented in the form
\be
\label{pool_sc}
G^{(m)}\stackrel{d}{=}\frac{1}{E+i\eta-\epsilon-\sum_{i=1}^{m} G_i^{(m)}}
\ee
and
\be
\label{pool_simple}
G \equiv G^{(m+1)}\stackrel{d}{=}\frac{1}{E+i\eta-\epsilon-\sum_{i=1}^{m+1} G_i^{(m)}},
\ee
where the symbol $\stackrel{d}{=}$ denotes the equality in distribution for random variables. On the right-hand-side of Eqs.~(\ref{pool_sc}) and (\ref{pool_simple}), 
$G_i^{(m)}$ are independent, identically distributed copies of the random variable $G^{(m)}$ and $\epsilon$ is a random variable with distribution 
$\gamma(\epsilon)$. 

Within the pool method, one finds numerically the distribution of $G^{(m)}$ as a fixed point of an iterative procedure based on the self-consistency equation, Eq.~(\ref{pool_sc}). Then the distribution of $G \equiv G^{(m+1)}$ is obtained from Eq.~(\ref{pool_simple}). After this, moments of the local density of states are easily evaluated:
\be
\left<\nu\right>_{BL} \equiv \nu = - \frac{1}{\pi}\lim_{\eta\to 0}\left<\Im G(E+i\eta)\right>
\ee
and 
\be
\left<\nu^2\right>_{BL}  = \frac{1}{\pi^2}\lim_{\eta\to 0}\left<[\Im G(E+i\eta)]^2\right>.
\ee
We have performed this procedure for disorder values  $W=8$, 10, and 12, and thus evaluated $\alpha(0)$ as given by Eq.~(\ref{alpha0-1}). In this calculation, we employed the pool size $M=2^{31}$ and the broadening $\eta=10^{-5}$, which are sufficient to reach the limiting behavior. Values of $\alpha(0)$ as obtained by the pool method are shown by star symbols in the left panel of Fig.~\ref{same}. At $W=8$ and $W=10$, a perfect agreement with results of the exact diagonalization is observed, which confirms the validity of our analysis. At $W=12$, the exact diagonalization results  give a somewhat smaller value of $\alpha(0)$ than that derived by the pool method. This is related to finite-size corrections in the exact diagonalization, which become more pronounced when the disorder approaches its critical value and thus the correlation volume $N_\xi$ increases. In other words, the system with $N=131072$ nodes is not large enough to accurately reach the asymptotic limit $N\to \infty$ for $W=12$. Indeed, as Fig. 3 of Ref. \onlinecite{tikhonov2016anderson} shows, at the system size $N=131072$ the value of $\alpha(0) N^2$ for this disorder is already quite close to its $N=\infty$ limiting value but is still slightly below it.

\subsection{Different wavefunctions}
\label{sec:different_wavefunctions}

\begin{figure*}[tbp]
\minipage{0.5\textwidth}\includegraphics[width=\textwidth]{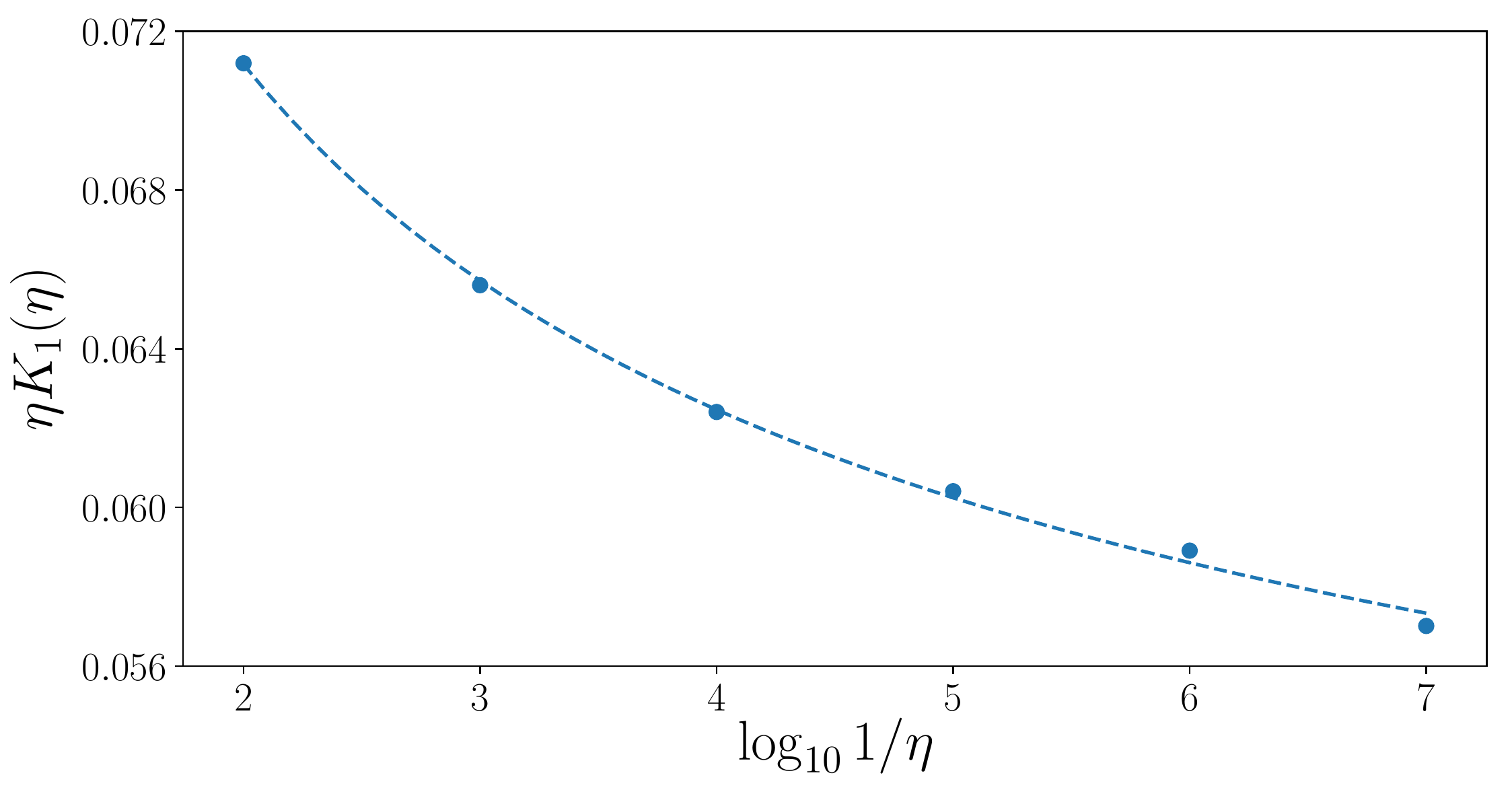}\endminipage
\minipage{0.5\textwidth}\includegraphics[width=\textwidth]{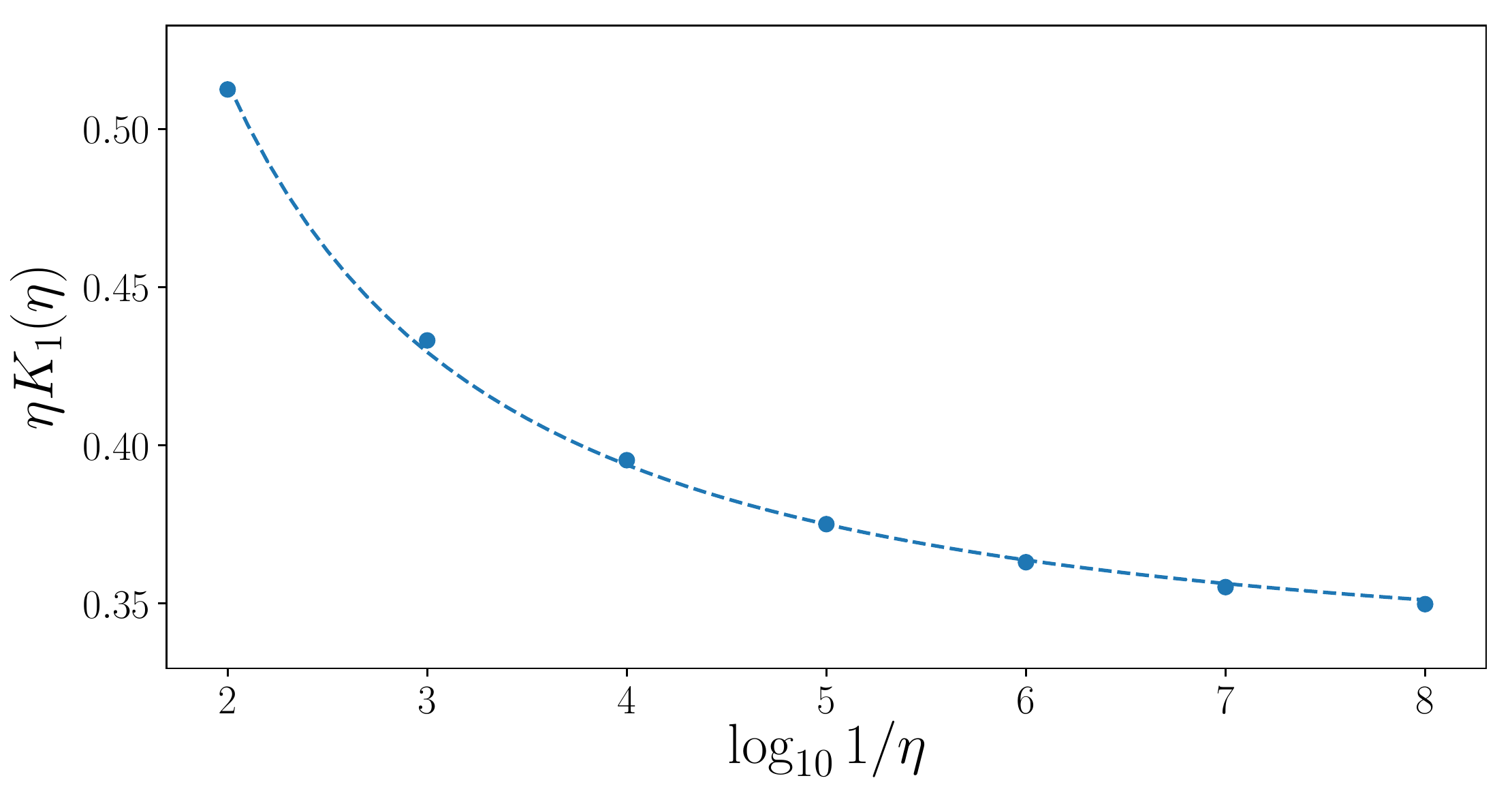}\endminipage
\caption{Same-point correlation function at imaginary frequency $K_1(\eta)$ at criticality. Left: $\eta K_1(\eta)$ as obtained by numerical solution of the self-consistency equations for the $n=1$ Anderson model via population dynamics (dots) and a fit to  $\eta K_1(\eta) = c_1^{(K)} + c_2^{(K)} \ln^{-\mu}(1/\eta)$ according to Eq.(\ref{K1-eta-with-subleading}), with $\mu = 1/2$ (dashed line), where $c_1^{(K)}\approx 0.041$. Right: analogous plot for the toy $\sigma$ model of Refs.~\cite{zirnbauer1990fourier,zirnbauer1991fourier,gruzberg96}, with a fit to Eq.~(\ref{K1-eta-with-subleading}) with $\mu=3/2$.}
\label{figmu}
\end{figure*}

We turn now to the analysis of the  correlation function $\beta(r,\omega)$ defined by Eq.~(\ref{sigmadef}). 
In analogy with Eq.~(\ref{def}), we use an identity relating $\beta(r,\omega)$ to averaged products of Green functions on RRG,
\be
\label{alphasigmaQ}
\alpha_{ij} \left(E\right) \delta \left(
\omega/\Delta\right) +\beta_{ij} \left(E,\omega \right) \bar{R
}\left( \omega \right) =\Delta^2 B_{ij}(E,\omega),
\ee
where 
\be
\bar{R}\left( \omega \right) =R\left( \omega \right) -\delta \left(\omega /\Delta \right)
\label{R-bar}
\ee
 is the non-singular part of the two-level correlation function (\ref{R-omega}) and 
\bea
\label{defB}
B_{ij}(E,\omega) &=& \left<\nu_{i}(E-\omega/2)\nu_{j}(E+\omega/2)\right> \nonumber \\
 &=& \frac{1}{2\pi^2}\Re\left[\left<G_R(i,i,E+\frac{\omega}{2})G_A(j,j,E-\frac{\omega}{2})\right.\right.\nonumber\\
&-& \left.\left.G_R(i,i,E+\frac{\omega}{2})G_R(j,j,E-\frac{\omega}{2})\right>\right]
\label{Bij}
\eea
is the correlation function of local densities of states at different energies and different spatial points.

We proceed with the evaluation of the correlation function (\ref{Bij}) for the $n=1$ RRG model in the same way as it was done in Sec.~\ref{s3.3}. In the delocalized phase and at $N \gg N_\xi$, we  obtain a zero-mode integral of the type (\ref{IPR-T-integral}), (\ref{Ug0T}), with the difference that it should be now evaluated at a real frequency $\omega$ and the real part should be taken. When this is done, the first $Q$-dependent structure in  Eq.~(\ref{Ug0T}) yields the random-matrix-theory (RMT) level correlation function, and the second one drops out. As a result, we get
\be 
B_{ij}(E,\omega) = \bar{R}_{\rm WD}(\omega) \frac{1}{2\pi^2}
\Re\left[K_1(r,\omega)-\left<G_R(0)\right>^2\right],
\label{Bij-result}
\ee
where $\bar{R}_{\rm WD}(\omega)$ is the Wigner-Dyson level correlation function and
\bea
\label{K1w}
K_1(r,\omega) = \langle G_R(i,i,E+\frac{\omega}{2}) G_A(j,j,E-\frac{\omega}{2}) \rangle_{\rm BL}.
\eea
We further use that the level correlation function $\bar{R}(\omega)$ on RRG is  given, in the delocalized phase and at $N \gg N_\xi$, by its RMT form $\bar{R}_{\rm WD}(\omega)$, see Sec.~\ref{s4}. Thus, Eqs.~(\ref{alphasigmaQ}) and (\ref{Bij-result}) yield
\be
\label{betamain}
\beta(r,\omega)=\frac{1}{2\pi^2 N^2}\Re\left[K_1(r,\omega)-\left<G_R(0)\right>^2\right].
\ee
For $r=0$ and for the SRM ensemble this result was obtained in Ref. \onlinecite{fyodorov1997strong}. 
Similarly to Eq. (\ref{alpha-r-deloc}), the formula (\ref{betamain}) expresses a correlation function of eigenfunctions on RRG with large $N$ in terms of a correlation function defined on an infinite Bethe lattice (or, equivalently, via a self-consistency equation). 

In the localized phase (or at criticality), the function integral of the type (\ref{gsuper}) for the evaluation of the average in Eq.~(\ref{alphasigmaQ}) is given by a single saddle point $g_0(x,y)$, which is now a solution of the self-consistency equation with finite $\omega$. The level correlation function $\bar{R}(\omega)$ is now equal to unity. As a result, we find again Eq.~(\ref{betamain})   which therefore has a very general validity. The condition on $N$ under which Eq.(\ref{betamain}) is valid at criticality is determined below, see Eq. (\ref{omegaN}) and text around it.

\subsubsection{The same point}
\label{subsubsec:samepoint}

Now we apply the general formula (\ref{betamain}) to specific regimes. We consider first the correlation function $\beta(0,\omega)$ at coinciding points, $r=0$. 
We begin this analysis by inspecting the behavior of $\beta(0,\omega)$ at criticality ($W= W_c$, or, more generally, $N_\xi \gg N$) and then turn to the delocalized phase ($W< W_c$ and $N \gg N_\xi$).  Performing in Eq.~(\ref{K-localized}) an analytical continuation to real frequency, $\eta\to - i\omega/2$, and setting $r=0$,
we get
\be
\label{K1loc}
K_1(r=0, \omega)\sim \frac{1}{-i\omega}.
\ee
Such a $1/\omega$ scaling of criticality would be in agreement with an expectation based on the extrapolation of finite-$d$ results to $d\to \infty$, see Eq.~(\ref{betasc}). However, according to Eq.~(\ref{betamain}), we have to take a real part of $K_1(0,\omega)$. The leading term  (\ref{K1loc}) therefore yields no contribution. We need thus to look for corrections to the leading behavior of the correlation function $K_1$ as obtained from the solution of the self-consistency equation, which is a rather challenging task. While we did not manage to evaluate analytically this correction, we can provide an ``educated guess'' for its form that is then verified and made more precise by a numerical solution of the self-consistency equation. Specifically, we expect that corrections at criticality are governed by inverse powers of $\ln1/\eta$:
\be
\label{K1-eta-with-subleading}
K_1(r=0) \simeq \frac{c_1^{(K)}}{\eta}+\frac{c_2^{(K)}}{\eta\ln^{\mu}1/\eta},
\ee
The emergence of logarithmic subleading factors from the solution of the self-consistency equation is very natural, since its kernel is nearly translational-invariant with respect to the variable $\ln x$ \cite{mirlin1991localization} or the analogous variable in the $\sigma$ model formalism\cite{efetov1985anderson,zirnbauer1986localization,zirnbauer1986anderson,efetov1987density,efetov1987anderson,verbaarschot1988graded}. In particular, the diffusion constant scales in the delocalized phase close to the transition point as $D \sim N_\xi^{-1} (\ln N_\xi)^3$.  The correlation volume $N_\xi$ sets a scale for the symmetry breaking in the delocalized phase; at criticality this role is played by $1/\eta$, which suggests the emergence of powers of $\ln 1/\eta$. We also note that a subleading factor in the form of a power of the logarithm of frequency is a natural counterpart of the subleading $r^{-3/2}$ factor in the $r$-dependence, see Eq.~(\ref{alphaloc}).

We have verified Eq.~(\ref{K1-eta-with-subleading}) by numerical solution of the self-consistency equation, both for the $n=1$ Anderson model and for the $\sigma$ model at the imaginary frequency $\eta$. The results for $\eta K_1(\eta)$ at criticality in the $n=1$ model are shown in the left panel of Fig.~\ref{figmu}; they fully confirm Eq.~(\ref{K1-eta-with-subleading}). The value of the exponent $\mu$ as determined numerically, is $\mu \simeq 1/2$. It seems likely that 1/2 represents an exact value of $\mu$ for this model;  it would be interesting to get this from the analytical solution of the self-consistency equation. 

The right panel of Fig.~\ref{figmu} presents the analogous results for the $\sigma$ model. Since the dependence on the compact variable $\lambda_2$ of the $\sigma$ model become irrelevant near criticality $W \simeq W_c$, one can reduce a self-consistency equation to that for a function of a single variable $\lambda_1$, which greatly improves the efficiency of the numerical analysis\cite{zirnbauer1986localization}. Further, we have replaced the corresponding kernel of the integral equation by that found for the hyperbolic superplane\cite{zirnbauer1990fourier,zirnbauer1991fourier,gruzberg96}---a ``toy version'' of the $\sigma$ model. 
All the essential properties of the kernel in this model are the same as for the reduced self-consistency equation of the proper $\sigma$ model but its functional form is much simpler. The obtained results fully confirm the expected behavior, Eq. (\ref{K1-eta-with-subleading}), which thus holds both for $n=1$ model as the $\sigma$ model. Interestingly, the numerical fit yields for the exponent of the logarithm $\mu \simeq 3/2$, i.e., a value different from that for the $n=1$ model. Again, it seems that 3/2 is an exact value of $\mu$ for the $\sigma$ model, and it would be interesting to obtain it analytically and to understand the source of the difference between both models in this aspect.

\begin{figure*}[tbp]
\minipage{0.5\textwidth}\includegraphics[width=\textwidth]{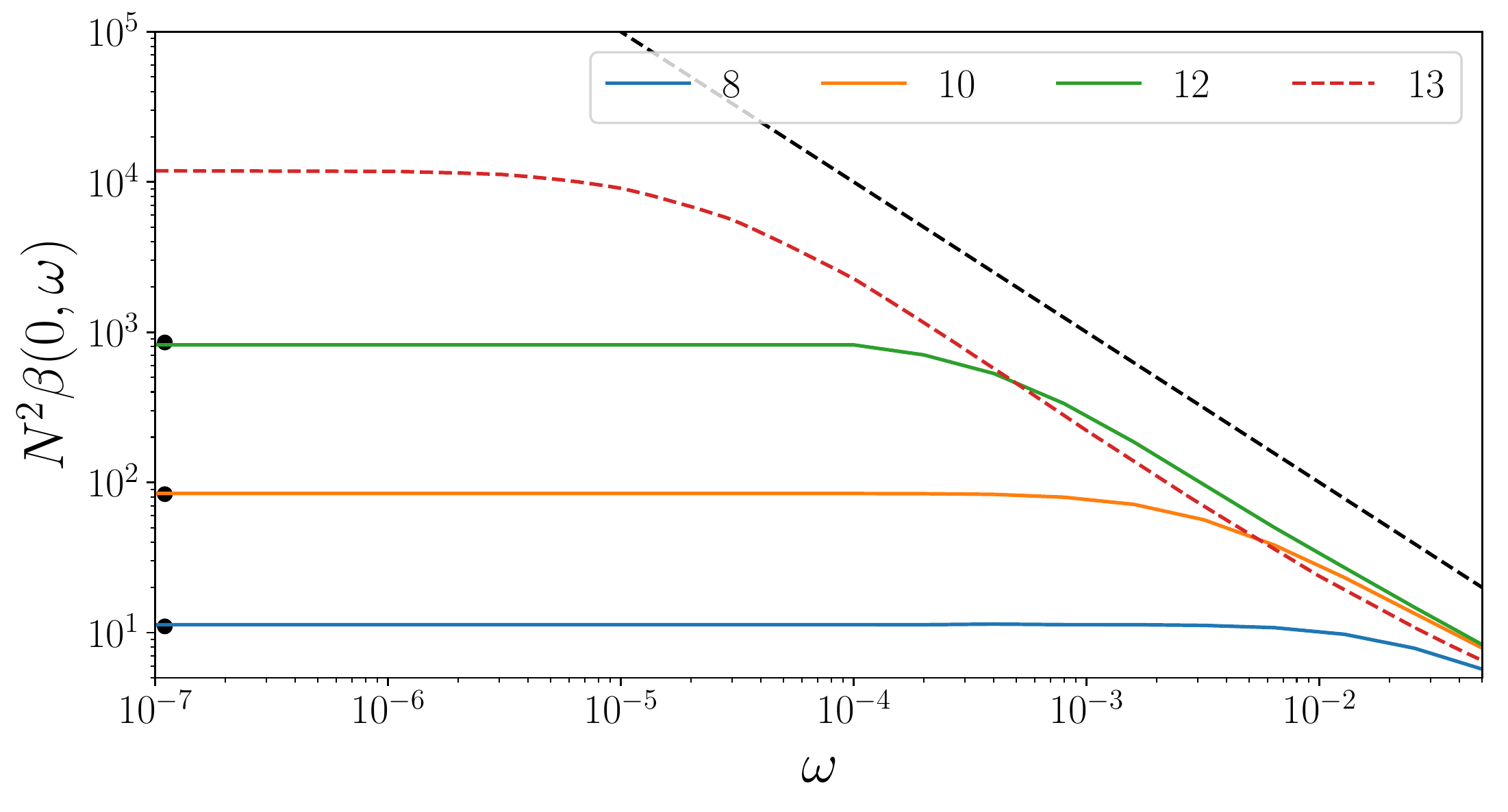}\endminipage
\minipage{0.5\textwidth}\includegraphics[width=\textwidth]{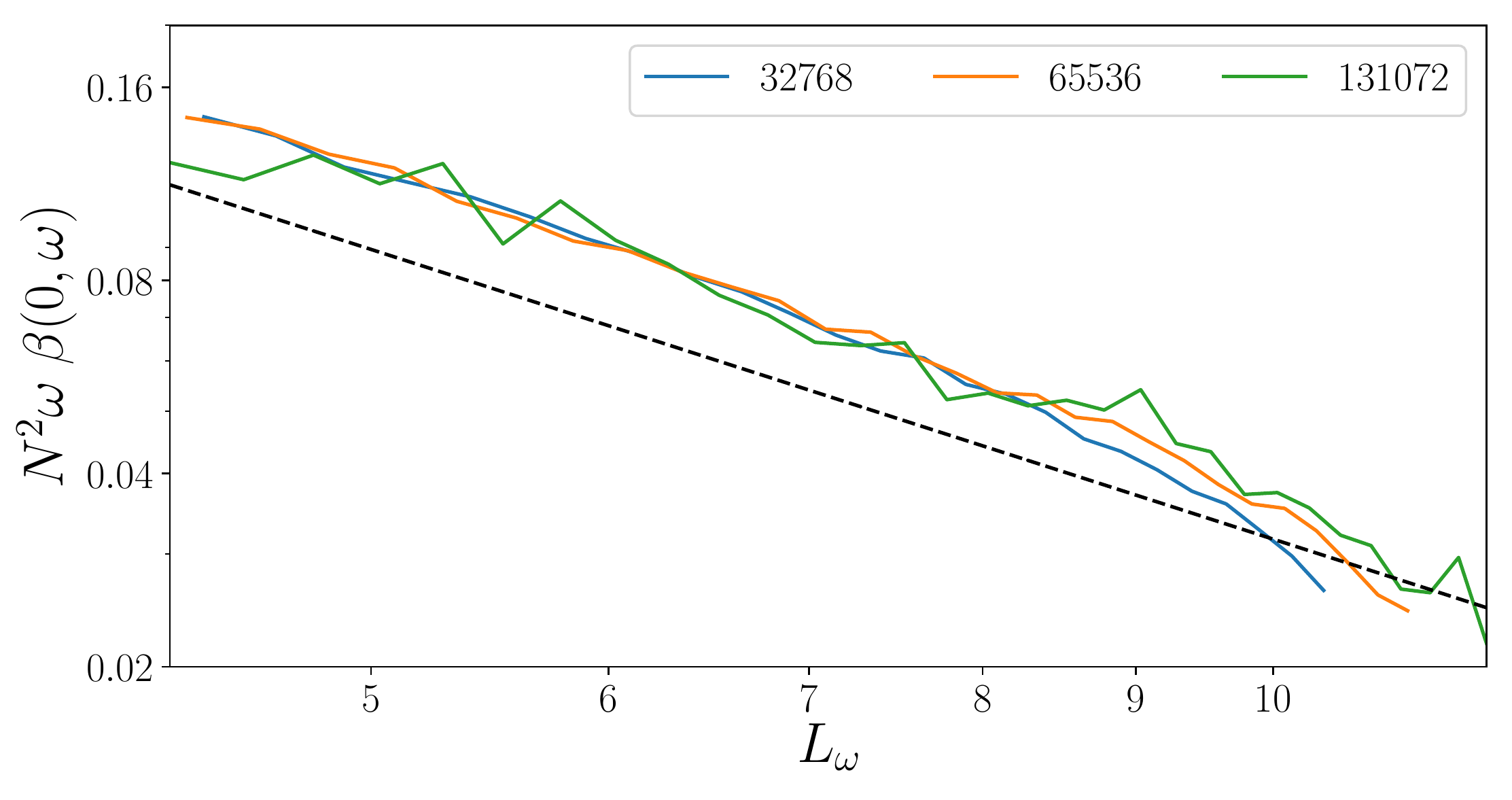}\endminipage
\caption{Correlation function $\beta(0,\omega)$ of different eigenfunctions at the same spatial point. Left: 
$N^2  \beta(0,\omega)$ as a function of frequency $\omega$ in the delocalized phase. Solid lines: exact diagonalization for system size $N=131072$, red dashed line: analytical result (\ref{betamain}) complemented by numerical solution of the self-consistency equations (\ref{1tra}), (\ref{2tra}), black dashed line: $1/\omega$ scaling. Black dots show $\alpha(0)/3$ for comparison. Right: 
$N^2 \omega \beta(0,\omega)$ as a function of $\ln(1/\omega)$ at the critical point for several system sizes (indicated in the legend). Dashed line: $1/\ln^{3/2}(1/\omega)$ scaling.}
\label{different}
\end{figure*}

It is instructive to compare the value of the IPR at criticality as obtained by exact diagonalization (see Sec.~\ref{s3.3} and Fig. \ref{p2loc}) with its value found from the solution of the self-consistency equation, see left panel of Fig. \ref{figmu}. According to Eq. (\ref{P2K1loc}):
\be
\label{iprloc}
\lim_{N\to\infty}P_2(N)=\frac{1}{\pi\nu}\lim_{\eta\to 0}\eta K_1(0,\eta).
\ee
It is natural to assume that finite-size corrections to the IPR at criticality have the same logarithmic form as corrections in Eq. (\ref{K1-eta-with-subleading}), with $\eta\to1/N$, which is a rational behind our fitting function (\ref{P2scalingloc}). The extrapolation of the exact diagonalization data shown in Fig. \ref{p2loc} yielded $\lim_{N\to\infty}P_2(N) \simeq 0.20$, see Sec.~\ref{s3.3}. On the other hand, the extrapolation of the population-dynamics results shown in the left panel of the Fig. \ref{figmu} yields $\lim_{\eta\to 0}\eta K_1(0,\eta) \simeq 0.041$ and thus $P_2 \simeq 0.24$ according to Eq. (\ref{iprloc}). This is in a reasonable agreement with the value obtained by exact diagonalization. Note that solution of the self-consistency equation provides a more accurate extrapolation since it allows one to proceed closer to the asymptotic limit. 

Performing in Eq.~(\ref{K1-eta-with-subleading}) the analytic continuation to a real frequency and substituting the result into Eq.~(\ref{betamain}), we find the following scaling of $\beta(0,\omega)$ at criticality:
\be
\label{betacrit}
\beta(0,\omega)\sim \frac{1}{N^2 \omega\ln^{\mu+1} 1/\omega}.
\ee
The applicability of the critical scaling (\ref{betacrit}) is limited, on the side of small $\omega$, by the finite size $N$ of the system. In the limit $\omega \to 0$ we have 
\be
\label{beta-crit-0}
\beta(0,\omega \to 0)\sim 1/N.
\ee 
To show this, we use the fact that $\beta(0,\omega \to 0) / \alpha(0) = 1/3$ in the delocalized phase at $N \gg N_\xi$, see below. By continuity, this implies that $\beta(0,\omega \to 0) / \alpha(0) \sim 1$ at criticiality, $N \ll N_\xi$. Using Eq.~(\ref{alphaloc}), we get Eq.~(\ref{beta-crit-0}). The matching between the low-frequency limiting value (\ref{beta-crit-0}) and the critical scaling (\ref{betacrit}) happens at a frequency 
\be 
\label{omegaN}
\omega_N\sim \frac{1}{N \ln^{\mu+1}N},
\ee
which is parametrically suppressed (by a logarithmic factor) compared to the mean level spacing $\Delta\sim 1/N$. 
To shed light on the physical significance of this logarithmic factor, we recall that, in a $d$-dimensional  system, the crossover between the corresponding two regimes happens at $\omega_N\sim g_*\Delta$, where $g_*$ is the critical value of conductance. The Anderson transition moves, with increasing $d$, further and further into the strong coupling regime, which means that $g_*$ decreases with increasing $d$, tending to zero in the $d \to \infty$ limit. (See Refs. \onlinecite{garcia2007dimensional,tarquini2017critical} for a systematic analysis of the evolution of properties of the Anderson transition with increasing $d$). Since RRG can be viewed as corresponding to the $d = \infty$ limit, it is expected that the constant $g_*$ is replaced by a function of $N$ that vanishes in the thermodynamic limit. 
This reduction of wave function correlations reflects itself also in reduction of level repulsion, as we discuss in Section \ref{s4} below.

We consider now the behavior of $\beta(0,\omega)$ in the delocalized phase, $W<W_c$ and $N\gg N_\xi$. In the small-frequency limit, we 
find, by comparing Eqs.~(\ref{betamain}) and (\ref{alpha-r-deloc}), 
\be
\label{beta-alpha}
\beta(0,\omega \to 0) = \frac{1}{3} \alpha(0) = \frac{N_\xi}{N^2}.
\ee
The factor $1/3$ in Eq.~(\ref{beta-alpha}) is the same as in the Gaussian ensemble of RMT.  Its emergence here is one more manifestation of the ergodicity of the delocalized phase on RRG.  The correlation function $K_1(0,\omega)$ and hence $\beta(0,\omega)$ remain nearly constant (i.e., frequency independent) for not too high frequencies.  At larger frequencies, the system enters the critical regime, so that Eq.~(\ref{betacrit}) holds. The crossover frequency $\omega_\xi$ can be estimated  by matching of Eqs.~(\ref{betacrit}) and (\ref{beta-alpha}); it is given by  $\omega_\xi\sim N_{\xi}^{-1}$, up to a logarithmic factor. To summarize, we get in the delocalized phase 
\be
\label{betascres}
\beta(0,\omega)\sim
\left\{
\begin{array}{cc} \displaystyle
 \frac{N_\xi}{N^2}, & \quad \omega<\omega_{\xi},\\[0.4cm]
 \displaystyle 
\frac{1}{N^2 \omega \ln^{\mu+1}1/\omega}, & \quad \omega>\omega_{\xi}.
\end{array}
\right.
\ee
This result largely confirms the expectation, Eq. (\ref{betasc}), based on the $d \to \infty$ extrapolation, improving it by an additional factor, weakly (logarithmically) dependent on frequency.

These predictions are fully supported by  numerical results for the correlation function $\beta(0,\omega)$ of different eigenfunctions at coinciding spatial points shown in Fig.~\ref{different}.  In the left panel, the results for the delocalized phase are presented. Solid lines (disorder $W=8,\;10,\;12$) are obtained by exact diagonalization of RRG model with the system size $N=131072$. We also show by dashed line in this plot results for 
$W=13$ as obtained  from Eq.~(\ref{betamain}) and (\ref{K1w}), with finite-$\omega$  correlations on an infinite Bethe lattice derived from the self-consistency equations (\ref{1tra}), (\ref{2tra}), see Sec.~\ref{sec:return_prob} below for technical details of this procedure. A very good agreement between both types of numerical results and the analytical prediction Eq. (\ref{betascres}) is clearly observed. Both the critical behavior ($1/\omega$, up to corrections that is difficult to observe in this plot) and the low-frequency saturation are evident. As an additional check, we show in this figure by dots the numerically obtained values of $\alpha(0)/3$; it is seen that 
$\beta(0,\omega)/\alpha(0)=1/3$, see Eq.~(\ref{beta-alpha}), is perfectly fulfilled. Numerical data for $\beta(0,\omega)$ shown in Fig.~\ref{different} provide two ways to extract the correlation length $\xi$:  from the value at $\omega \to 0$ and from the crossover scale in the $\omega$ dependence. Both of them yield values of $\xi(W)$ close to those shown in the inset of Fig.~\ref{same}. 

 In the right panel of Fig.~\ref{different}, results of exact diagonalization of the RRG model at the critical point are shown. To verify the subleading logarithmic factor in the critical behavior, Eq.~(\ref{betascres}), we plot here $N^2 \beta(0,\omega)$ multiplied by $\omega$ as a function of $\ln(1/\omega)$. The results fully confirm the power-law scaling with respect to $\ln(1/\omega)$, and the corresponding exponent is perfectly consistent with the value $\mu=1/2$ as derived from our result (\ref{betamain}) with the help of numerical solution of the self-consistency equation at an imaginary frequency $\omega\to2i\eta$.

\subsubsection{Return probability}
\label{sec:return_prob}

\begin{figure*}[tbp]
\minipage{0.5\textwidth}\includegraphics[width=\textwidth]{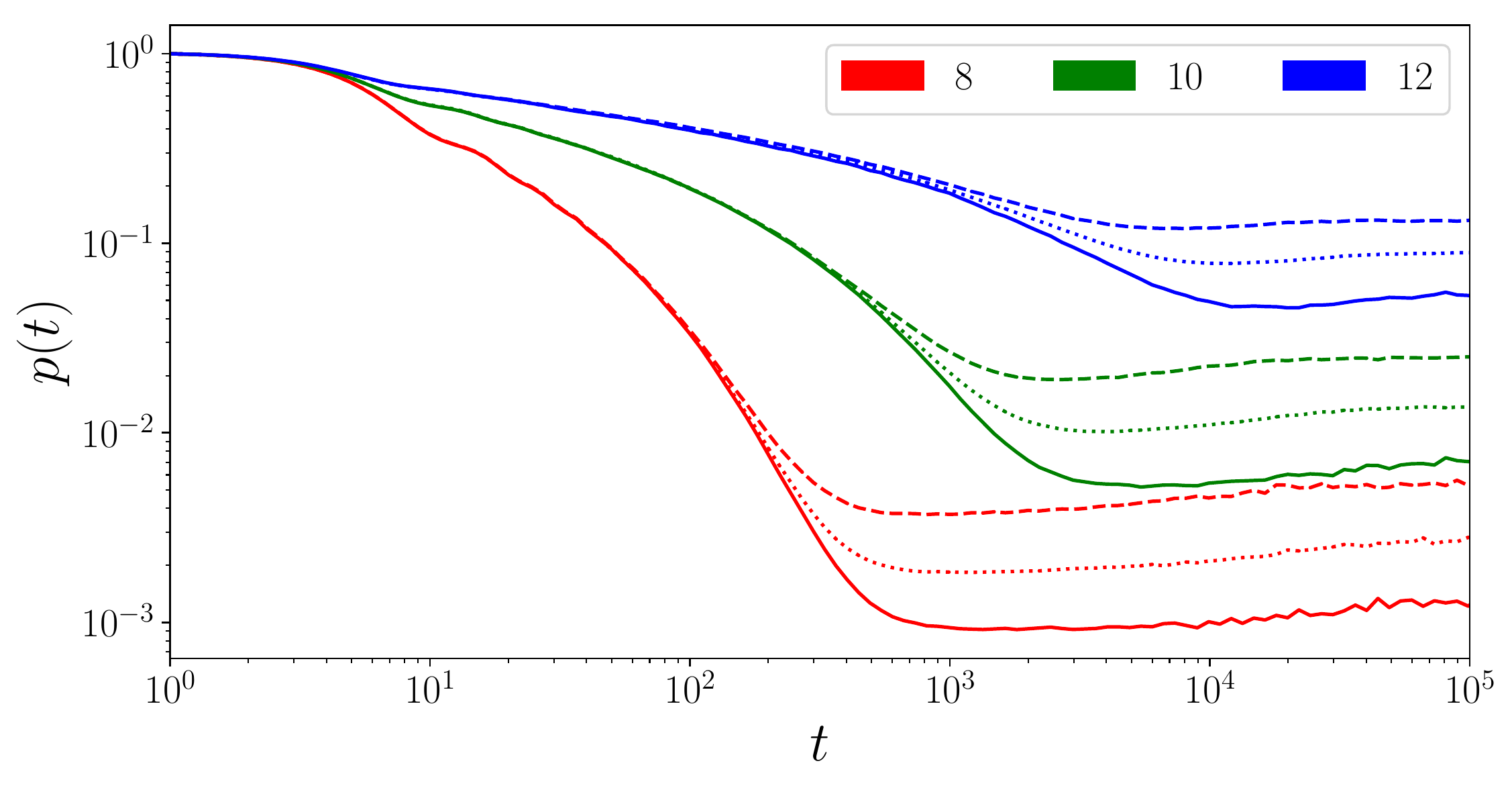}\endminipage
\minipage{0.5\textwidth}\includegraphics[width=\textwidth]{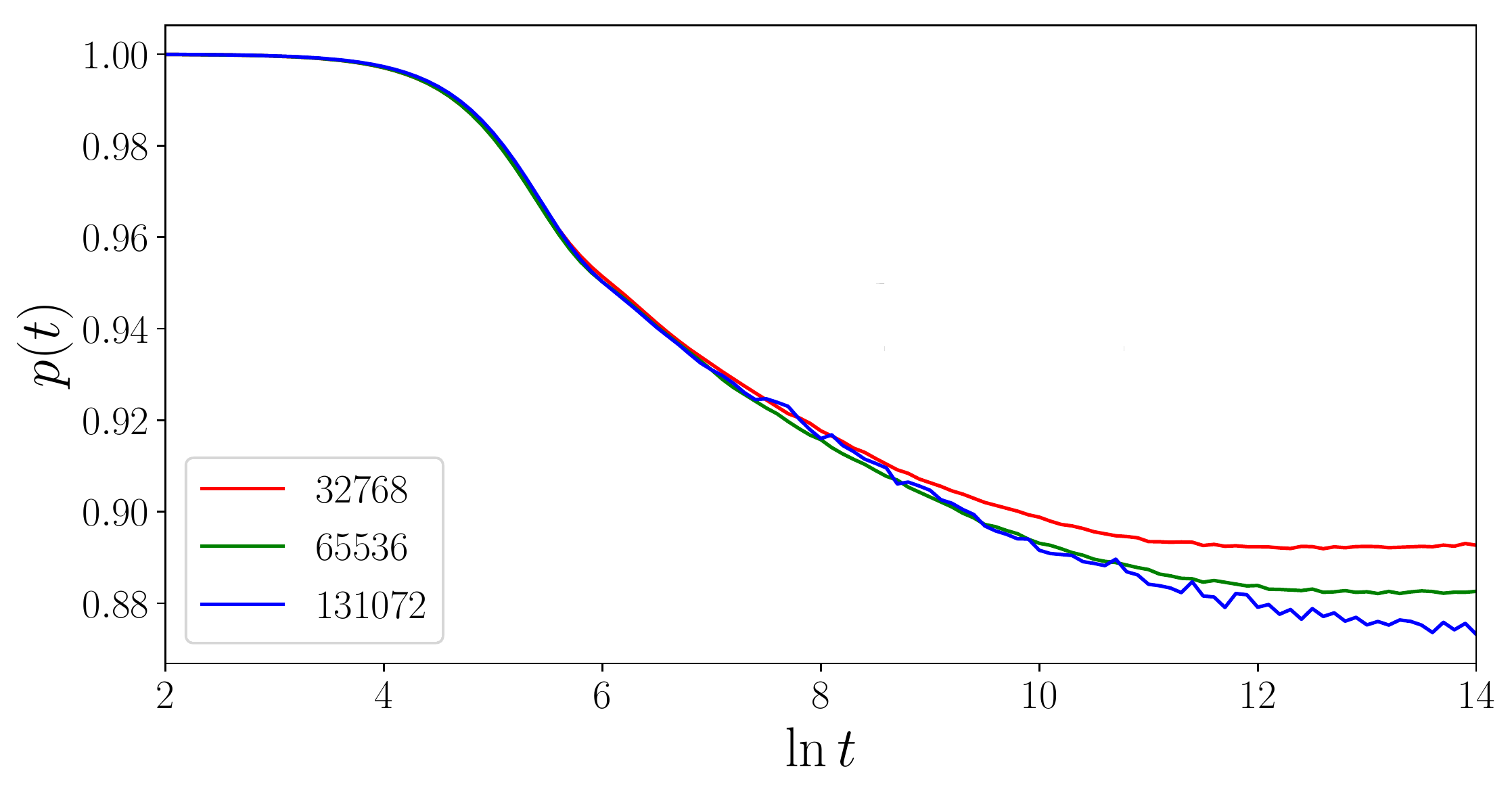}\endminipage
\caption{Return probability $p(t)$ on RRG as obtained by exact diagonalization, with a projection on states near the band center (1/16 of all states). Left: delocalized phase with disorder $W=8$ (red), 10 (green), and 12 (blue). The dashed, dotted, and solid lines correspond to system sites $N=32768$, $65536$, and $131072$, respectively. A crossover from fast decay to finite-size saturation at $p_\infty = N \alpha(0) \sim N_{\xi}/N$ is observed. Right: critical point for several system sizes, illustrating a slow  approach to $p_\infty$, see Eq.~(\ref{pcrit}).}
\label{returna}
\end{figure*}

The correlation function $\beta(0,\omega)$ has the meaning of the return probability in the Fourier domain. Indeed, consider spreading of a state, localized at $t=0$ at a given site $j$ and evaluate probability $p(t)$ to find it at the same site at a later time $t>0$. Formally, it is defined as follows (averaging over the initial site $j$ is performed):
\be
p(t)=\left<\frac{1}{N}\sum_{j} |\bra{j}e^{-i\hat H t}\ket{j}|^2\right>.
\label{pt}
\ee
Performing the Fourier transformation, $p(\omega)=\int e^{i\omega t}p(t)dt$, and expanding in eigenfunctions $| \alpha\rangle$, one gets
\be
\label{Kdef}
p(\omega)=\left<\frac{1}{N}\sum_{\alpha\beta}\sum_j |\bra{\alpha}j\rangle|^2 |\bra{\beta}j\rangle|^2\delta(\omega+E_\alpha-E_\beta)\right>.
\ee
Using the spectral representation (\ref{Kdef}), we can straightforwardly express the function $p(\omega)$ in terms of eigenfunction correlators defined in Eqs. (\ref{alphadef}) and (\ref{sigmadef}):
\bea
\label{pomega}
p(\omega) &=& N\delta(\omega)\int dE\: \nu(E)\alpha_E(0)  \nonumber \\
&+& N^2 \int dE\: \nu^2(E) R_E(\omega) \beta_E(0,\omega).
\eea
Thus, while the $t\to\infty$ limit of the return probability is related to $\alpha_E(0)$, its time dependence is encoded in $\beta_E(0,\omega)$.

As defined by Eqs.~(\ref{pt}), (\ref{Kdef}),  the return probability is a quantity integrated over the whole band of energies $E$. States with very different energies will also have very different localization (delocalization) properties. One thus has to define a return probability for a given energy $E$ (or for its relatively narrow vicinity). This can be done by projecting the state $|j\rangle$ in Eq.~(\ref{pt}) on a subspace of eigenstates with energy around $E$,
\bea
p_E(t) & = & \tilde p_E(t) / \tilde p_E (0);  \label{pEt-normalized}\\
\tilde p_E(t) & = &  \left<\frac{1}{N}\sum_{j}  |\bra{j} \hat P_E \: e^{-i\hat H t} \: \hat P_E \ket{j}|^2\right>,
\label{pEt}
\eea
where $\hat P_E$ is the corresponding projector. Equation (\ref{pEt-normalized}) simply ensures the normalization $p_E(0)=1$. 
Equation (\ref{pEt}) yields a spectral representation for $\tilde p_E(\omega)$ of the same form (\ref{Kdef}), but now with summation restricted to the states $|\alpha\rangle$, $|\beta\rangle$ with energies close to $E$.  The long-time behavior of $p_E(t)$ defined in this way will be determined by the functions $\alpha_E(0)$ and $\beta_E(0,\omega)$ at the chosen energy $E$.  Below we assume that such a projection is performed and omit the subscript $E$ in the notation for the return probability; i.e., $p(t)$ below means $p_E(t)$. 

We now proceed by analyzing the behavior of the return probability $p(t)$ first at criticality and then in the delocalized phase. 
In the critical regime, both contributions in Eq. (\ref{pomega}) scale with $N$ in the same way, reaching a finite value in the limit of $N\to\infty$. Fourier transforming the asymptotics in Eq. (\ref{betacrit}) we find:
\be
\label{pcrit}
p(t) \simeq p_{\infty} + \frac{c^{(P)}}{\ln^{\mu}t}, \qquad t \to \infty,
\ee
were $p_{\infty} \sim N \alpha(0) \sim 1$.

In the metallic regime ($W<W_c$ and $N \gg N_\xi$), one expects the return probability $p(t)$ to be described by a classical random walk over the tree, characterized by a diffusion coefficient $D$. 
For such a random walk, the probability that a particle will be found at the starting point after time $t$ is given by\cite{monthus1996random}
\be
\label{pdiff}
p(t)\sim \frac{1}{(Dt)^{3/2}}e^{-Dt}.
\ee
It is known that the diffusion coefficient in the $n=1$ model and in the $\sigma$ model on the Bethe lattice  behaves as 
\be
\label{diff}
D \sim N_\xi^{-1} \ln^3 N_ \xi \ ,
\ee
where $N_\xi$ scales according to Eq.~(\ref{Nxi}). 
The diffusive decay of $p(t)$ corresponds to the  second contribution in Eq. (\ref{pomega}). Once the particle spreads roughly uniformly over the graph, the decay saturates at a value given by the first term in Eq. (\ref{pomega}),
\be
p_\infty \sim N \alpha(0) \sim \frac{N_\xi}{N}.
\label{p-infty-deloc}
\ee

In Fig.~\ref{returna}, we show results for return probability $p(t)$ as obtained by exact diagonalization of the RRG model. 
We used the definition (\ref{pEt-normalized}), (\ref{pEt}), with $\hat P_E$ projecting on $1/16\textrm{th}$ of all eigenstates around the energy $E=0$. The left panel of Fig.~\ref{returna} shows the results for delocalized regime and the right panel for the critical point.  We discuss the latter data first. The exact-diagonalization results for the critical point confirm our prediction (\ref{pcrit}) based on the analytical treatment supplemented by numerical solution of the self-consistency equation. The return probability approaches logarithmically slowly its limiting value $p_\infty$.

In the left panel of Fig.~\ref{returna} the data for disorder $W=8$, 10, and 12 are shown, in each case for three different system sizes. For $W=8$, we observe a fast drop of $p(t)$, with a strongly pronounced curvature in the log-log plot, which indicates an exponential character of decay, in line with the expectation (\ref{pdiff}). At long times, this decay is limited by the system size $N$, and $p(t)$ saturates at the value $p_\infty$, in full consistency with the prediction (\ref{p-infty-deloc}). A similar behavior is observed for $W=10$.  For still stronger disorder, $W=12$, the return probability exhibits a nearly flat part up to the time $t \sim 10^{-3}$. This is related to the fact that $W=12$ is already sufficiently close to the critical point, so that a nearly-critical behavior is observed in the  intermediate range of times.  For longer times, the clear downward curvature appears, indicating a crossover to an exponentially fast decay characteristic for the delocalized regime. However, it does not have much time to develop, in view of the saturation dictated by the system size. 

In order to verify the manifestation  of the prediction  (\ref{betamain}) in the time domain and to find $p(t)$ for longer times, we have evaluated $\beta(0,\omega)$ by population dynamics (such data for disorder $W=13$ have been already shown in the left panel of Fig.~\ref{different}) and then Fourier-transformed it to get $p(t)$ in the limit $N \to \infty$. To calculate $\beta(0,\omega)$,  as given in terms of infinite-Bethe lattice correlation function by Eqs.~(\ref{betamain}), (\ref{K1w}), 
we used self-consistency equations for the joint distribution function of two Green functions on different energies, $u=G_R(i,i,E+\omega/2)$ and $v=G_A(i,i,E-\omega/2)$, which provide a natural generalization of Eqs. (\ref{fm}) and (\ref{fm1}):
\begin{widetext}
\begin{eqnarray}
f^{(m)}(u,v) &=& \int d\epsilon \: \gamma(\epsilon)\int \left( \prod_{r=1}^{m} du_r \,  dv_r \, f^{(m)}(u_r,v_r)  \right) \nonumber \\
& \times & \delta\left[ u - \frac{1}{E+\frac{\omega}{2}+i\eta -\epsilon - \sum_{r=1}^{m} {u_r}} \right]    
\delta\left[ v - \frac{1}{E-\frac{\omega}{2}-i\eta -\epsilon  - \sum_{r=1}^{m}{v_r}} \right]  ;
\label{2tra}  \\
f^{(m+1)}(u,v) &=& \int d\epsilon \: \gamma(\epsilon) \int \left( \prod_{r=1}^{m+1} du_r \,  dv_r \, f^{(m)}(u_r,v_r) \right)  \nonumber  \\
& \times &  \delta\left[ u - \frac{1}{E+\frac{\omega}{2}+i\eta -\epsilon - \sum_{r=1}^{m+1} {u_r}} \right]    
\delta\left[ v - \frac{1}{E- \frac{\omega}{2}-i\eta -\epsilon  - \sum_{r=1}^{m+1}{v_r}} \right].
\label{1tra}
\end{eqnarray}
\end{widetext}
In a recent work\cite{PhysRevLett.117.104101,metz2017level}, these equations emerged within a saddle-point approach for the replica action in course of the analysis of the level number variance on RRG.

\begin{figure}[tbp]
\includegraphics[width=0.5\textwidth]{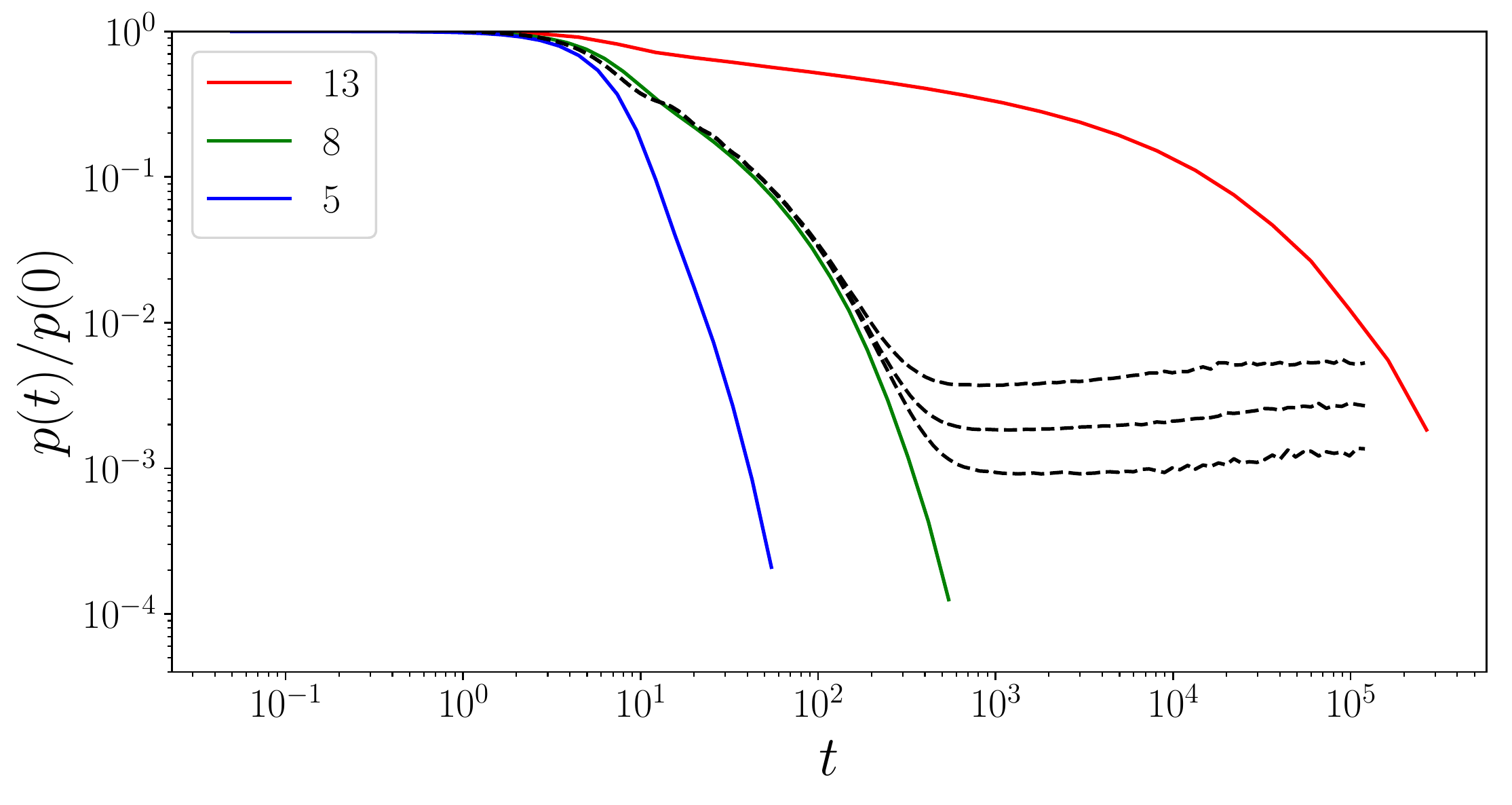}
\caption{Return probability $p(t)$ in the delocalized phase obtained as a Fourier-transform of $\beta(0,\omega)$ found via solution of the self-consistency equation (full lines; blue, green, and red colors correspond to disorder $W=5$, 8, and 13, respectively.)  Black dashed lines: return probability $p(t)$  for $W=8$ and three different system sizes, $N=32768$, 65536, and 131072, as found by exact diagonalization (the same curves as in the left panel of Fig.~\ref{returna}).}
\label{returng}
\end{figure}

We have solved the self-consistency equations (\ref{2tra}), (\ref{1tra}) by using the pool size $M=10^{10}$ and the broadening $\eta=10^{-7}$ and determined $\beta(0,\omega)$ according to Eqs.~(\ref{betamain}), (\ref{K1w}). Next, we have Fourier-transformed this quantity to evaluate $p(t)$. In order to stay in full correspondence with the exact-diagonalization study, we have limited the frequency integration to the $1/16\textrm{th}$  of the energy band, with a Gaussian smoothing. (This modified the short-time behavior but  had essentially no influence on the long-time behavior of the resulting $p(t)$, up to a constant prefactor.)    
The return probability $p(t)$ obtained in this way from the numerical solution of the self-consistency equation is shown in Fig.~\ref{returng} for disorder strength $W=5$, 8, and 13.  For comparison, we also show in this figure the exact-diagonalization results for $W=8$ from Fig.~\ref{returna}. A very good agreement is observed, which confirms once more the validity of our saddle-point analysis, yielding RRG observables at $N \gg 1$ in terms of infinite-Bethe-lattice correlation functions expressed in terms of the solution of the self-consistency equation. Contrary to the RRG results, the population-dynamics data do not show saturation at long $t$. This is because in the population-dynamics analysis we have not included the term proportional to $\alpha(0)$ in Eq.~(\ref{pomega}), thus effectively calculating $p(t)$ in the $N \to \infty$ limit. Of course, a numerical approach does not allow us to proceed to arbitrarily long times. An obvious limitation is that we can reliably calculate $p(t)$ only as long it is much larger than the inverse pool size $M^{-1}$. In fact, there exists a considerably more stringent restriction related to the fact that $p(t)$ becomes exponentially small at long times and is represented by a sum of many terms that are not particularly small. Thus, the numerical evaluation of $p(t)$ involves huge cancellations, requiring a high precision of intermediate computations. This limits (via statistical fluctuations due to finitenes of the pool size) the smallest $p(t)$ until which we were able to proceed reliably with the population-dynamics calculation shown in Fig.~\ref{returng}.  

\begin{figure*}[tbp]
\minipage{0.5\textwidth}\includegraphics[width=\textwidth]{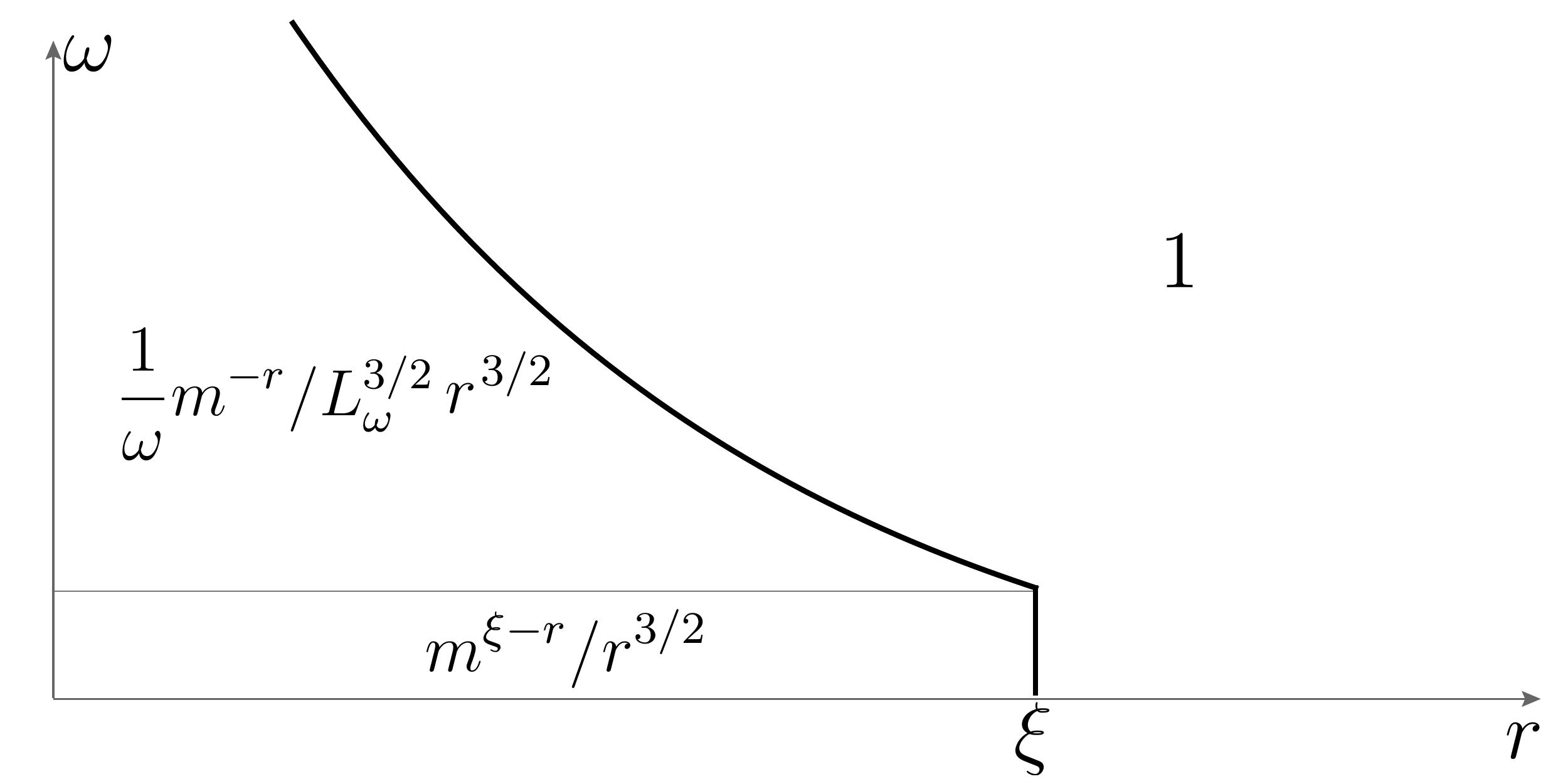}\endminipage
\minipage{0.5\textwidth}\includegraphics[width=\textwidth]{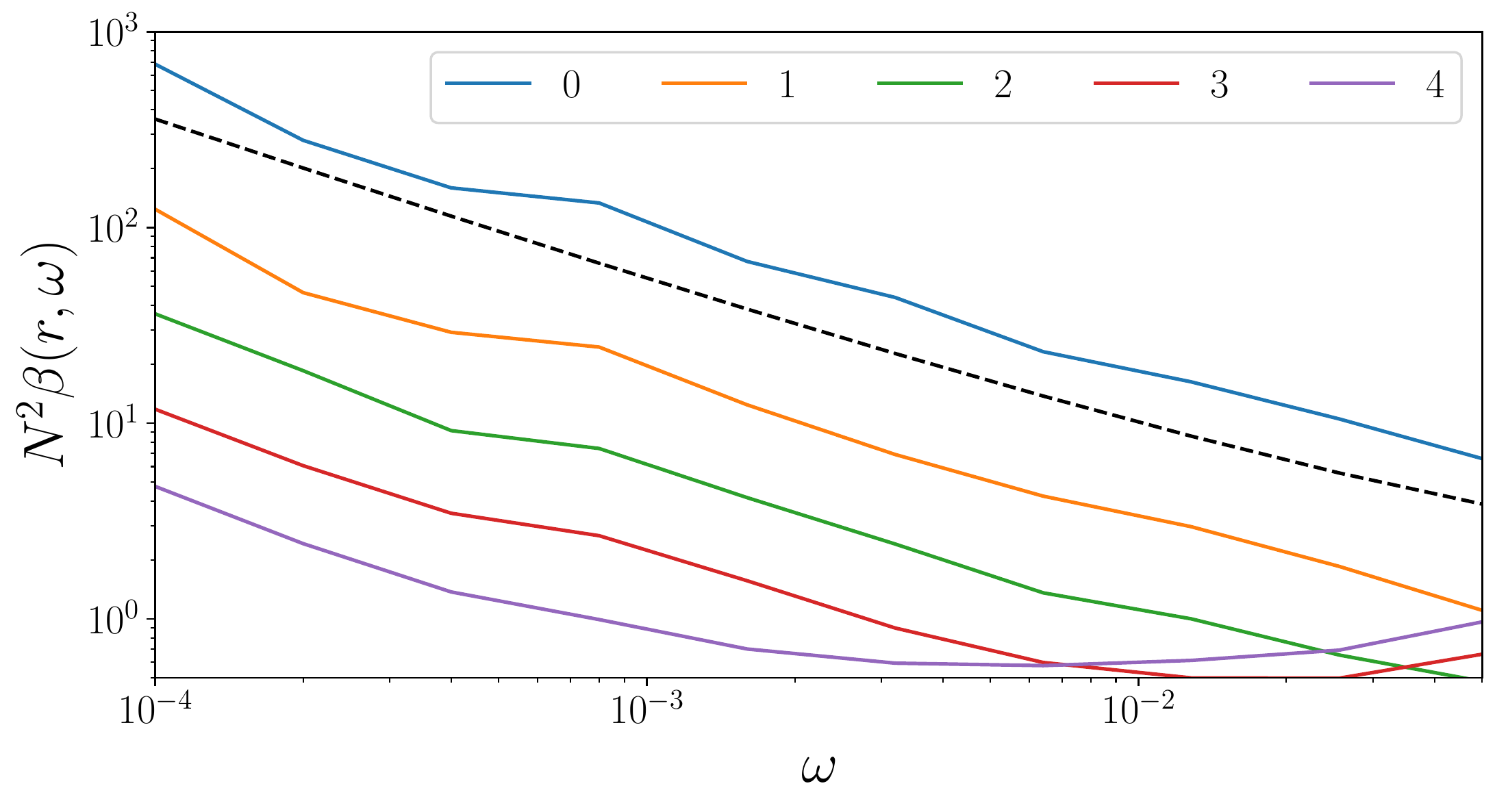}\endminipage
\caption{Correlation function $\beta(r,\omega)$ of different wavefunctions at different spatial points. Left: scaling behavior of $N^2\beta(r,\omega)$ in different part of the $(r,\omega)$ plane. The lines are $L_\omega = \xi$ (thin) and  $r = \min(L_\omega,\xi)$ (thick), up to logarithmic corrections. Here  $L_\omega$ is the length corresponding to the frequency $\omega$ at criticality, Eq.~(\ref{L-omega}).  
Right: results of exact diagonalization at the critical point ($W=18$) for the system size $N=32768$. Dashed line: predicted $\omega$-dependence $1/\omega L^{3/2}_\omega$, see Eq.~(\ref{betaresr}).
}
\label{sigmar}
\end{figure*}

It should be mentioned that the return probability on RRG was studied numerically in two recent works, Refs.~\onlinecite{biroli2017delocalized} and \onlinecite{bera18}, with the conclusions that $p(t)$ behaves as a power law of time\cite{biroli2017delocalized} and as a stretched exponential\cite{bera18} in the delocalized phase. We believe that these conclusions resulting from fits in a limited interval of time do not have a fundamental meaning but rather reflect crossover regions between the logarithmically slow critical behavior and the exponentially fast asymptotic decay. We have emphasized above that reaching numerically the actual long-$t$ asymptotics of $p(t)$ is a very difficult task. 

\subsubsection{Different points}

Finally, let us discuss the correlation function $\beta(r,\omega)$ at different spatial points ($r \gg 1$) for $W < W_c$.  It is given in terms of the infinite-Bethe-lattice correlation function by Eq.~(\ref{betamain}). For sufficiently low frequencies, $\omega < \omega_\xi$, the correlation function $K_1(r,\omega)$ is essentially frequency-independent and is given by Eq.~(\ref{K1-deloc}). As a result, we find that in this regime $\beta(r,\omega)$ is essentially equal to $\alpha(r)$ and scales with $r$ according to Eq.~(\ref{alphadeloc}). 

At criticality (high-frequency range), a fully analytical calculation on the basis of Eq.~(\ref{betamain}) encounters the same difficulty as for $\beta(0,\omega)$:  one has to find a subleading correction to $K_1(r,\omega)$ that survives when one takes the real part. We thus resort to an ``educated guess'' based on the results obtained above, which we then verify by exact diagonalization. We expect that at criticality, where $\beta(r,\omega)$ should have a strong dependence on both $\omega$ and $r$, these dependencies decouple, in analogy 
with the $d$-dimensional case, Eq.~(\ref{betasc}). Then the $\omega$-dependence is  given by the second line of Eq.~(\ref{betascres}). As to the $r$-dependence, it is expected to be the same as at low frequencies, $\omega < \omega_\xi$, as otherwise there would be a mismatch between the two regimes.
Indeed, the critical regime should cross over to the metallic one at the frequency 
$\omega_\xi$, as for $r=0$, see Eq.~(\ref{betascres}). The condition $\omega \sim \omega_\xi$ can be equivalently written (up to logarithmic correction) as $L_\omega \simeq \xi$, where 
\be
\label{L-omega}
L_\omega = \frac {\ln (1/\omega)} {\ln m} 
\ee
is the length corresponding to frequency $\omega$ at criticality. Finally, at sufficiently large $r$, the correlation function $\beta(r,\omega)$ is given by the disconnected part of Eq.~(\ref{betamain}) that yields $1/N^2$. 
 Combining everything, we obtain
\be
\label{betaresr}
\beta(r,\omega)\sim
\left\{
\begin{array}{ll}
 \displaystyle \frac {m^{\xi-r}}{N^{2}r^{3/2}}, & \quad \omega < \omega_{\xi} \quad \textrm{and} \quad r<\xi;\\[0.4cm]
 \displaystyle \frac{m^{-r}}{N^2 \omega  L_{\omega}^{3/2}r^{3/2}}, & \quad \omega_{\xi}<\omega<m^{-r}.
\end{array}
\right.
\ee
The first line of Eq.~(\ref{betaresr}) correspond to the ``metallic'' regime with the hierarchy of the length scales $r < \xi < L_\omega$, while the second line to the critical regime with $r < L_\omega < \xi$, in analogy with Eq.~(\ref{betasc}). (When writing these conditions, we discard logarithmic corrections, cf.  Eq.~(\ref{alphadeloc}) and the footnote after it\cite{foot2}.)  If $r > \min (L_\omega, \xi)$, the correlation function decouples, 
$N^2\beta(r,\omega)\simeq 1$. 

These analytical results and expectations are summarized in the left panel of Fig.~\ref{sigmar} where the predicted behavior of $\beta(r,\omega)$ is shown in the form of a diagram in the $(r,\omega)$ plane. In this diagram, the thick line $r=\min(\xi,L_{\omega})$ bounds the region in the $(r,\omega)$ plane, where correlation function $\beta(r,\omega)$ is dominated by its connected part. This domain is further split into two parts by a thin line $L_{\omega}=\xi$, in correspondence with Eq.~(\ref{betaresr}).  

These predictions are fully supported by results of exact diagonalization of an RRG system at criticality shown in the right panel of Fig.~\ref{sigmar}.  It is clearly seen in this figure, that dependencies $\beta(r,\omega)$ form, on the logarithmic scale, a set of parallel curves at different values of $r$. This means a factorization of $r$ and $\omega$ dependencies in the critical regime, which was the assumption that we used to obtain the second line of Eq.~(\ref{betaresr}). 

Equation (\ref{betaresr}) shows that our expectations based on an extrapolation of finite-$d$ results, Eq.~(\ref{betasc}), were correct from the point of view of leading terms (exponential in $r$ and power-law in $\omega$). On the other hand, Eq.~(\ref{betaresr}) makes these results more precise, as it contains also subleading factors (power-law in $r$ and logarithmic in $\omega$). These factors, although subleading, are quite important. In particular, they provide a convergence of $r$-summation ($\omega$-integration) at small $r$ (respectively, $\omega$) in the sum rules:
\bea
\beta(0,\omega) + \sum_{r=1}^\infty (m+1)m^{r-1}\beta(r,\omega) &=& 1/N; \\
\nu \int d\omega \: \beta(0,\omega) &=& 1/N, 
\eea
when applied to the critical regime. This convergence is a necessary consequence of the scaling $\alpha(0) \sim \beta(0,\omega \to 0) \sim 1/N$ at criticality. This should be contrasted to a critical system at any finite $d$, for which the $r$-summation ($\omega$-integration) is controlled by large $r$ (respectively, $\omega$) and $\alpha(0) \sim \beta(0,\omega \to 0)$ is additionally suppressed by a fractional power of the system size. Also, we note that the subleading, logarithmic-in-$\omega$ factor controls the logarithmically slow time dependence  of $p(t)$ at criticality, see Sec.~\ref{sec:return_prob}.


\section{Spectral correlations}
\label{s4}

\subsection{From finite-$d$ models to RRG}
\label{s4.1}

Eigenenergies $E_k$ of disordered tight-binding models on $d$-dimensional lattices are correlated random variables whose statistics has been investigated intensively for several decades. Two most popular means to characterize the multivariate distribution function of energy levels $\mathcal{P}(\{E_i\})$ are the statistics of $P(\omega)$ of spacings between adjacent levels  and the two-level correlation function $R(\omega)$, Eq.~(\ref{R-omega}). An
important role in studies of the level statistics is played by the universal Wigner-Dyson distribution characterizing a Gaussian ensemble of the random matrix theory. This statistics  is characterized by repulsion between adjacent levels that has,  in the case of orthogonal symmetry [Gaussian orthogonal ensemble (GOE)], the form
$$ P(\omega\ll\Delta) \simeq \frac{\pi \omega}{2\Delta}.$$ 
The non-singular part (\ref{R-bar}) of the GOE  level correlation function reads\cite{mehta2004random,efetov1983kb} 
\begin{eqnarray}
\label{rmtgoe}
\bar{R}_{\rm WD}(\omega) &=& 1 -\frac{\sin ^{2}\pi s}{\left(\pi s\right) ^{2}} 
- \left[\frac{\pi}{2}\mathrm{sgn}(s)-\mathrm{Si}(\pi s)\right] \nonumber \\
& \times& \left[\frac{\cos\pi s}{\pi s}-\frac{\sin \pi s}{(\pi s)^2}\right], \ \ \ s = \frac{\omega}{\Delta}.
\end{eqnarray}

There is a close correspondence between the two-level correlation function $R(\omega)$ and the variance $\Sigma_2(\omega)$ of the number of levels $I(\omega)$ within a band of the width $\omega$,
\be
\label{variance}
\Sigma_2(\omega)=\left< I^{2}(\omega )\right>-\left<I(\omega )\right>^{2}.
\ee
Specifically, $\Sigma_2(\omega)$ can be expressed in terms of $R(\omega)$ as follows:
\be
\label{varianceR}
\Sigma_2(\omega)=\frac{2}{\Delta^2}\int_0^\omega(\omega-\omega^\prime)\left[R(\omega^\prime)-1\right]d\omega^\prime.
\ee
The level number variance is a convenient characteristics of the rigidity of the spectrum at relatively large energy scales, $\omega \gg \Delta$. 
In particular, for the Poisson and GOE statistics the level number variance reads 
\begin{eqnarray}
& \Sigma_2(\omega)  = \omega/\Delta, & \qquad \text{Poisson}, \\[0.1cm]
& \displaystyle \Sigma_2(\omega) \simeq \frac{2}{\pi^2}\ln\frac{2\pi\omega}{\Delta}, & \qquad \text{GOE}; \ \omega \gg \Delta.
\end{eqnarray}
Another closely related measure of the level fluctuations is the spectral compressibility
\be
\label{kappadef}
 \kappa(\omega)=\Delta \frac{d\Sigma_2(\omega)}{d\omega}=\frac{2}{\Delta}\int_0^{\omega}\left[R(\omega^\prime)-1\right] d\omega^\prime,
\ee
with the following behavior in the two limits
\begin{eqnarray}
& \kappa(\omega)  = 1, & \qquad \text{Poisson},  \label{kappa-P} \\[0.1cm]
& \displaystyle \kappa(\omega) \simeq \frac{2}{\pi^2}\frac{\Delta}{\omega}, & \qquad \text{GOE}; \ \omega \gg \Delta. \label{kappa-WD}
\end{eqnarray}
Numerically, it is advantageous to evaluate a slightly different quantity
\be
\label{chi-omega}
\chi(\omega)=\Sigma_2(\omega)/\left< I(\omega )\right>.
\ee
For the Poisson statistics it is identical to $\kappa$ (i.e., equal to unity), Eq.~(\ref{kappa-P});  in the Wigner-Dyson case it differs from $\kappa$, Eq.~(\ref{kappa-WD}), by a logarithmic factor only,
\be
\displaystyle \chi(\omega) \simeq \frac{2}{\pi^2}\frac{\Delta}{\omega} \ln\frac{2\pi\omega}{\Delta},  \qquad \text{GOE}; \ \omega \gg \Delta. 
\label{chi-WD}
\ee

A review of the behavior of $R(\omega)$ and $\Sigma_2(\omega)$ in a metallic sample of spatial dimensionality $d < 4$ can be found in Refs.~\cite{aronov1995fluctuations,mirlin00}; we briefly outline the key results that will be important for the analysis below. 

For frequencies  below the Thouless energy 
$E_{\rm Th} \sim D/L^2$ (which is much larger that the level spacing $\Delta$ for a metallic sample) the Wigner-Dyson statistics applies, implying, in particular,
Eq.~(\ref{rmtgoe}) for $R(\omega)$ and Eq.~(\ref{chi-WD}) for $\chi(\omega)$.
The Thouless energy has a physical meaning of the inverse time of diffusion through the sample.

Above the Thouless energy, the connected level correlation function $R^{(c)}(\omega) = R(\omega) - 1$ is dominated by the term $R^{(c)}_{\rm diff}(\omega)$ originating from diffusive  modes\cite{altshuler1986repulsion}:
\be
R^{(c)}_{\rm diff}(\omega)  =\frac{\Delta ^{2}}{\pi ^{2}}\Re\sum_{q \ne 0}\frac{1}{\left( Dq^{2}-i\omega \right) ^{2}}.
\label{R-diff}
\ee
For $d < 4$ the sum is controlled by the infrared limit, $q \sim \pi/L$, yielding
\be
\label{ASd}
R^{(c)}(\omega)\sim g^{-d/2}\left(\frac{\omega}{\Delta}\right)^{d/2-2}, 
\ee
where $g \sim E_{\rm Th} /\Delta$ is the dimensionless conductance. The relative level number variance $\chi(\omega)$ is then given by
\be
\label{chi-AS}
\chi(\omega)\sim g^{-d/2}\left(\frac{\omega}{\Delta}\right)^{d/2-1}.
\ee
We are particularly interested in the situation, when the system is close to the Anderson transition, on its metallic side, so that the correlation length $\xi$ is large but the system size is still larger, $L \gg \xi$. In this case, the diffusive regime, see Eqs. (\ref{ASd}), (\ref{chi-AS}) extends up to the frequency given by the Thouless energy $E_{\rm Th}(\xi) \sim D/\xi^2$ of a sample of the size $\xi$, which is of the order of the level spacing $\Delta_\xi$ of such a sample.

For higher frequencies  the system enters the critical regime. The level correlation function in this regime was studied in Ref. \onlinecite{aronov1995spectral}, with the result
\be
\label{KLAA}
R^{(c)}(\omega) \sim (\Delta_{\xi}/\Delta)^{1-\gamma}\left(\omega/\Delta\right)^{-2+\gamma},
\ee 
where $\gamma=1-1/\nu d$ and $\nu$ is the correlation length exponent (see also numerical study Ref. \onlinecite{braun1995spectral}). The spectral compressibility $\kappa$ as well as the relative level number variance $\chi$ are given by a non-trivial constant in the critical regime: $\kappa(\omega), \chi(\omega) \simeq \chi_*$ with $0 < \chi_* < 1$\cite{aronov1995fluctuations}. (This behavior was first proposed on a basis of numerical simulations in Ref.~\cite{altshuler1988repulsion}.) The critical ``spectral compressibility'' $\chi_*$ is a universal characteristics of the Anderson transition, i.e., it depends only on dimensionality and on symmetry class but not on the sample geometry and on microscopic details of the model. 
Corrections to $\chi_*$ are parametrically small and follow from Eq.~(\ref{KLAA}):
\be
\chi(\omega) = \chi_* + \delta \chi(\omega); \qquad \delta \chi(\omega) \sim (\Delta_{\xi}/\omega)^{1-\gamma}.
\label{chi-omega-crit}
\ee

Finally, the behavior of $R(\omega)$ and $\chi(\omega)$ at highest frequencies depends  on short-scale details of the microscopic model. In Ref.~\cite{aronov1995fluctuations} it was determined for an $n$-orbital model in $d$ dimensions. We will not need it for our discussion below.

After this reminder of previous results for $d$-dimensional systems with $d < 4$, we turn to the levels statistics on RRG.  Proceeding in the same way as for the eigenfunction correlations in Sec.~\ref{s3}, we will first formulated conjectures based on the extrapolation of finite-$d$ results to $d\to \infty$. After this, we will 
substantiate this conjectures by analytical calculations using the field-theoretical approach presented in Sec.~\ref{s3.2}. Finally, we will complement the analytical results by numerical exact-diagonalization study. 

Thus, we first extend the results for the above three regimes of the levels statistics (RMT, diffusive, and critical) to $d \to \infty$, with an expectation that this will correctly describe RRG. The RMT regime is fully universal once the frequency is normalized to the level spacing. Turning to the diffusive regime, we observe that the sum in Eq.~(\ref{R-diff}) diverges in the ultraviolet for $d > 4$.   This means that the sum should be cut off at largest $q$ and that the frequency $\omega$ in the denominator can be discarded. The result is 
\be
R^{(c)}_{\rm diff}(\omega) \simeq a \Delta,
\label{R-diff-large-d}
\ee
and thus 
\be
\chi (\omega)  \simeq a \omega.
\label{chi-diff-large-d}
\ee
If the sum over $q$ in Eq.~(\ref{R-diff}) is cut off at the inverse mean free path $l^{-1}$ (which is the largest momentum for which the diffusive approximation is still meaningful), one gets the following result for the coefficient  $c$ in these formulas:
\be
a \sim \frac{l^{4-d}}{\nu D^2}.
\label{c-diff}
\ee
In fact, the value of this coefficient is expected to be dependent on ultraviolet details of the model, since momenta $q \gg \l^{-1}$ will in general also contribute. This does not affect, however, the scaling of the level statistics (\ref{R-diff-large-d}) and  (\ref{chi-diff-large-d}) with $\Delta$ and $\omega$. We thus conclude that Eqs.~(\ref{R-diff-large-d}) and  (\ref{chi-diff-large-d}), with a coefficient $a$ that is independent on the system size and on frequency (i.e., depends only on short-scale details of the model and on degree of disorder), should be generically valid in $d > 4$ dimensions. We will continue calling this regime ``diffusive'' although the sum in Eq. (\ref{R-diff}) is of ultraviolet character.  It is a natural conjecture that Eqs. (\ref{R-diff-large-d}), (\ref{chi-diff-large-d}) are also valid on RRG. 

Finally, we discuss the critical regime. The critical spectral compressibility $\kappa_*$ is expected to tend to its Poisson value unity in the limit $d \to \infty$. Further, the exponent $1- \gamma = 1/\nu d$ governing the frequency-depending correction $\delta\chi(\omega)$  in Eq.~(\ref{chi-omega-crit}) tends to zero in the limit $d \to \infty$. Our previous experience (Sec.~\ref{s3}) tells us that this likely implies a logarithmic dependence of $\delta\chi$ on $\omega$. 

Having summarized our expectations based on finite-$d$ results, we can proceed with a direct analytical and numerical studies of the level statistics on RRG.

\subsection{Level number fluctuations on RRG:  Field-theoretical results and numerical simulations}

By definition, the RRG two-level correlation function can be written as
\be
\label{R-omega-B}
R(\omega)= \Delta^2 \sum_{ij} B_{ij}(E, \omega),
\ee
where $B_{ij}(E,\omega)$ is the correlation function of local densities of states defined by Eq.~(\ref{defB}). The correlation function $B_{ij}(E,\omega)$ was calculated within the supersymmetry approach in Sec.~\ref{sec:different_wavefunctions}. 
Using the result Eq. (\ref{Bij-result}), we get for the connected part $R^{(c)}(\omega)$ of the level correlation function:
\be
\label{RWDDiff}
R^{(c)}(\omega) = R^{(c)}_{\rm WD}(\omega)+R^{(c)}_{\textrm{diff}}(\omega),
\ee
where 
\be
\label{Rdiff}
R_{\textrm{diff}}^{(c)}(\omega) = \frac{\Delta}{2\pi^2 \nu} \sum_r m^r\Re K_{1}^{(c)}(r,\omega)
\ee
and $K_{1}^{(c)}(r,\omega) = K_1(r,\omega)-\left|\left<G_R(0)\right>\right|^2$ is the connected part of the Bethe-lattice correlation function (\ref{K1w}).  For $\omega \ll N_\xi^{-1}$ the correlation function $K_{1}^{(c)}(r,\omega)$ is essentially independent of $\omega$ and thus can be replaced by $K_{1}^{(c)}(r,0) \equiv K_{1}^{(c)}(r) \simeq K_2(r)$. Using Eq.~(\ref{K2-deloc}), we see that the sum over $r$ in Eq.~(\ref{Rdiff}) converges at $r \sim 1$,  yielding
\be
\label{RRes}
R_{\textrm{diff}}^{(c)}(\omega)\sim\frac{N_\xi}{N}, \qquad   \omega<N_{\xi}^{-1}.
\ee
The corresponding behavior of the relative number variance $\chi(\omega)$ is
\be
\label{chi-RRG-diff}
\chi(\omega) \sim  N_\xi\, \omega.
\ee

Let us analyze the obtained result Eq. (\ref{RWDDiff}). Since the second term, Eq. (\ref{RRes}) is proportional to $1/N$, the Wigner-Dyson statistics is valid up to a relatively large (in comparison to level spacing $\Delta$) scale. This  is another manifestation of ergodicity of the RRG ensemble in the whole delocalized phase $W < W_c$.  Similar to other manifestations of ergodicity, such as  the $1/N$ scaling of the IPR, the RMT level statistics sets in when the system is sufficiently large, $N \gg N_\xi$. The second term in Eq.~(\ref{RWDDiff}) as given by Eqs.~(\ref{Rdiff}), (\ref{RRes}) is fully analogous to the diffusive contribution
(\ref{R-diff}), (\ref{R-diff-large-d}) in $d > 4$ dimensions:  it is independent of $\omega$, proportional to $\Delta$, and controlled by short spatial scales.  

In order to determine the frequency scale at which the level correlation function loses its universal (RMT) character, we compare the two contributions in Eq.~(\ref{RWDDiff}). Using $R^{(c)}_{\rm WD}(\omega) \sim (\Delta/\omega)^2$ and Eq.~(\ref{RRes}), we find the crossover scale
\be
\omega_c\sim \frac{1}{ (N N_\xi)^{1/2}}.
\label{omega-c}
\ee
This frequency replaces the Thouless energy as the characteristic scale for departure from universal (RMT) behavior. It is worth emphasizing, however, that the  $\omega_c$ is much smaller than the inverse time required for a diffusing particle to reach the boundary of the system (which usually serves as a definition of Thouless energy). Indeed, the latter time scales only logarithmically with the volume $N$. 

We turn now to the discussion of the critical regime. We have seen in Sec.~\ref{subsubsec:samepoint} that the correlations between critical eigenstates are fully developed only up to the scale $\omega_N$, Eq.~(\ref{omegaN}), which is suppressed in comparison to the level spacing $\Delta$ by a logarithmic factor $\ln^{3/2}N \gg 1$. This means that the region of strong level repulsion experiences analogous suppression: the level repulsion is strong only for $\omega \lesssim \omega_N$. Therefore, with increasing $N$, the  level statistics of the RRG model at criticality approaches the Poisson statistics, as also confirmed by exact diagonalization. One of manifestation of this behavior is the Poisson value of the critical level compressibility, 
\be
\chi_* = 1.
\label{chi-star}
\ee
This should be contrasted to a finite-$d$ system at Anderson transition, for which the critical statistics is intermediate between Wigner-Dyson and Poisson forms \cite{shklovskii1993statistics} and, consequently, $0 < \chi_* < 1$.  In view of the logarithmic dependence of $\omega_N / \Delta$ on $N$, the approach to the critical value (\ref{chi-star}) is expected to be logarithmically slow in frequency,  in agreement with the argument (based on the $d\to \infty$ limit) presented in the end of Sec.~\ref{s4.1}. 

\begin{figure}[tbp]
\includegraphics[width=0.5\textwidth]{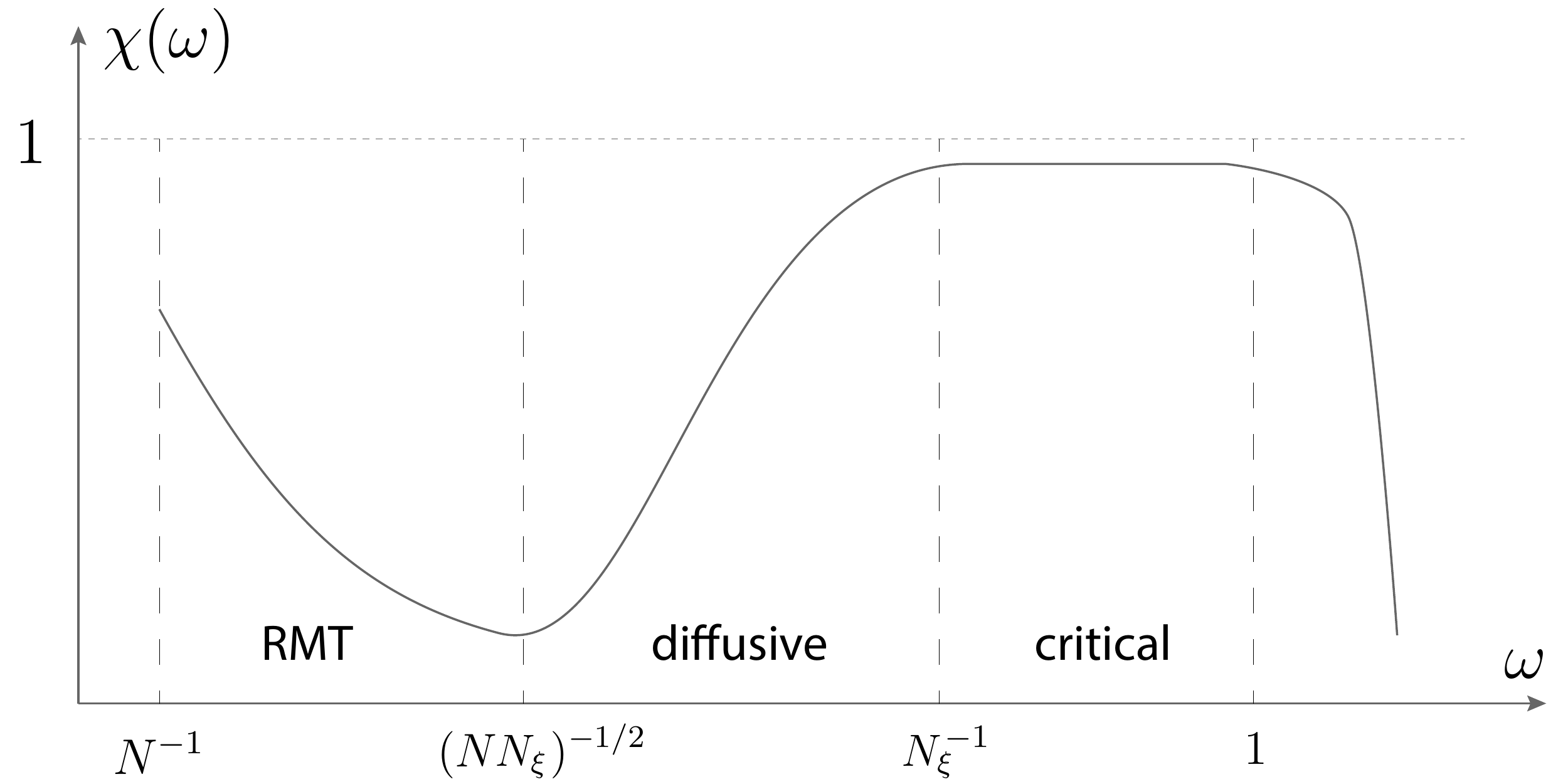}
\caption{Schematic representation of analytical predictions for the spectral statistics: Relative level number variance $\chi(\omega)$ in the metallic phase, $W < W_c$ close to the transition point, $N \gg N_\xi \gg 1$.  The behavior of $\chi(\omega)$ in the RMT regime is given by Eq.~(\ref{chi-WD}), in the diffusive regime by Eq.~(\ref{chi-RRG-diff}), and in the critical regime by Eq.~(\ref{chi-star}) with a logarithmic correction.
}
\label{ls}
\end{figure}

Our analytical results for the relative level number variance  $\chi(\omega)$ for a delocalized RRG system close to the metal-insulator transition, $N\gg N_{\xi}\gg 1$, are summarized in Fig.~\ref{ls}.

\begin{figure}[tbp]
\includegraphics[width=0.5\textwidth]{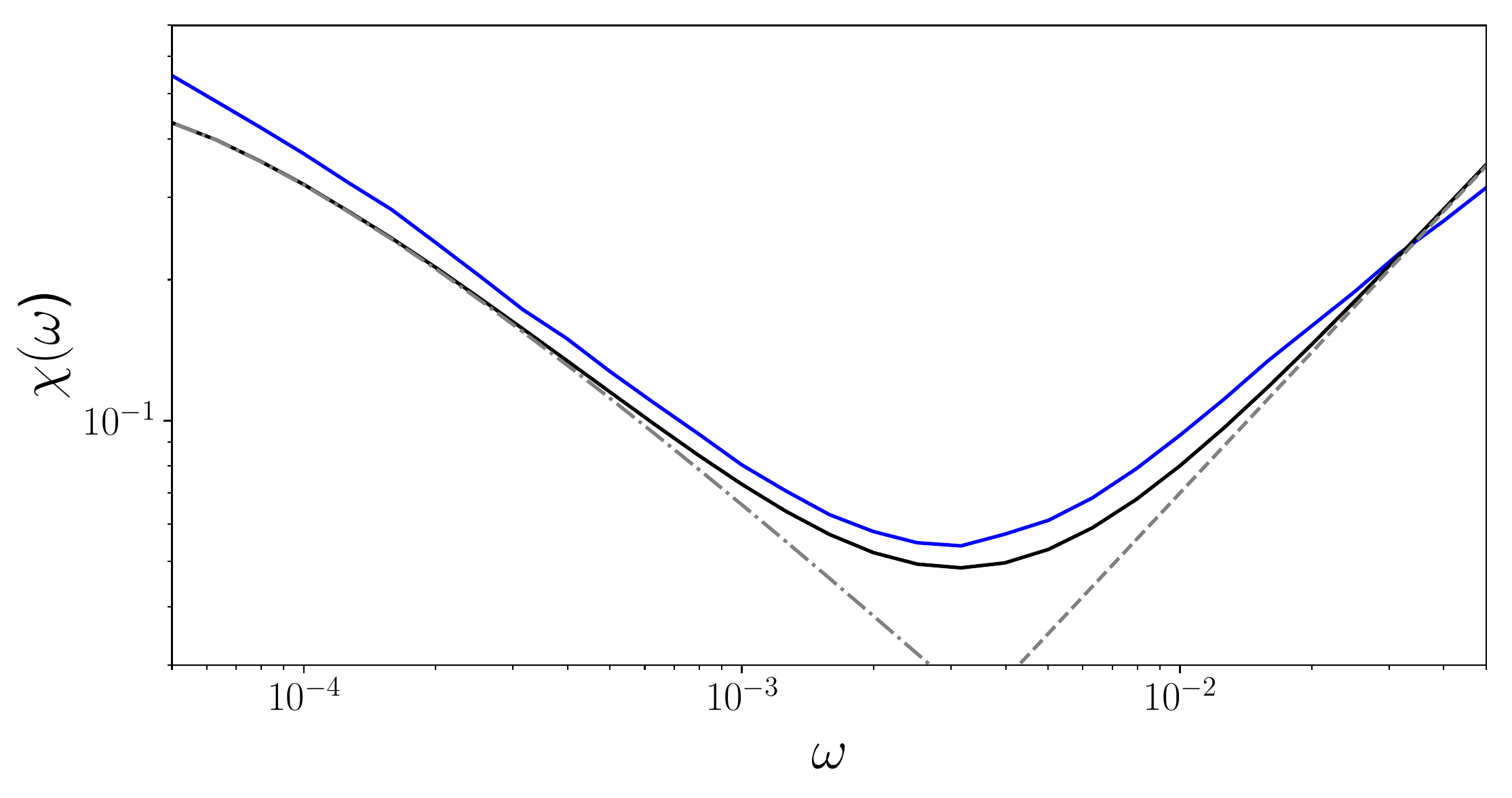}
\caption{Relative level number variance $\chi(\omega)$ as obtained by exact diagonalization. System with $N=131072$ sites, disorder $W=8$. 
Blue full line: exact diagonalization, black full line: analytical prediction (\ref{RWDDiff}), which is a sum of the RMT contribution, Eq.~(\ref{chi-WD}), shown by  dash-dotted line and the diffusive contribution (\ref{chi-RRG-diff}) shown by dashed line.
}
\label{figvar1}
\end{figure}

\begin{figure}[tbp]
\includegraphics[width=0.5\textwidth]{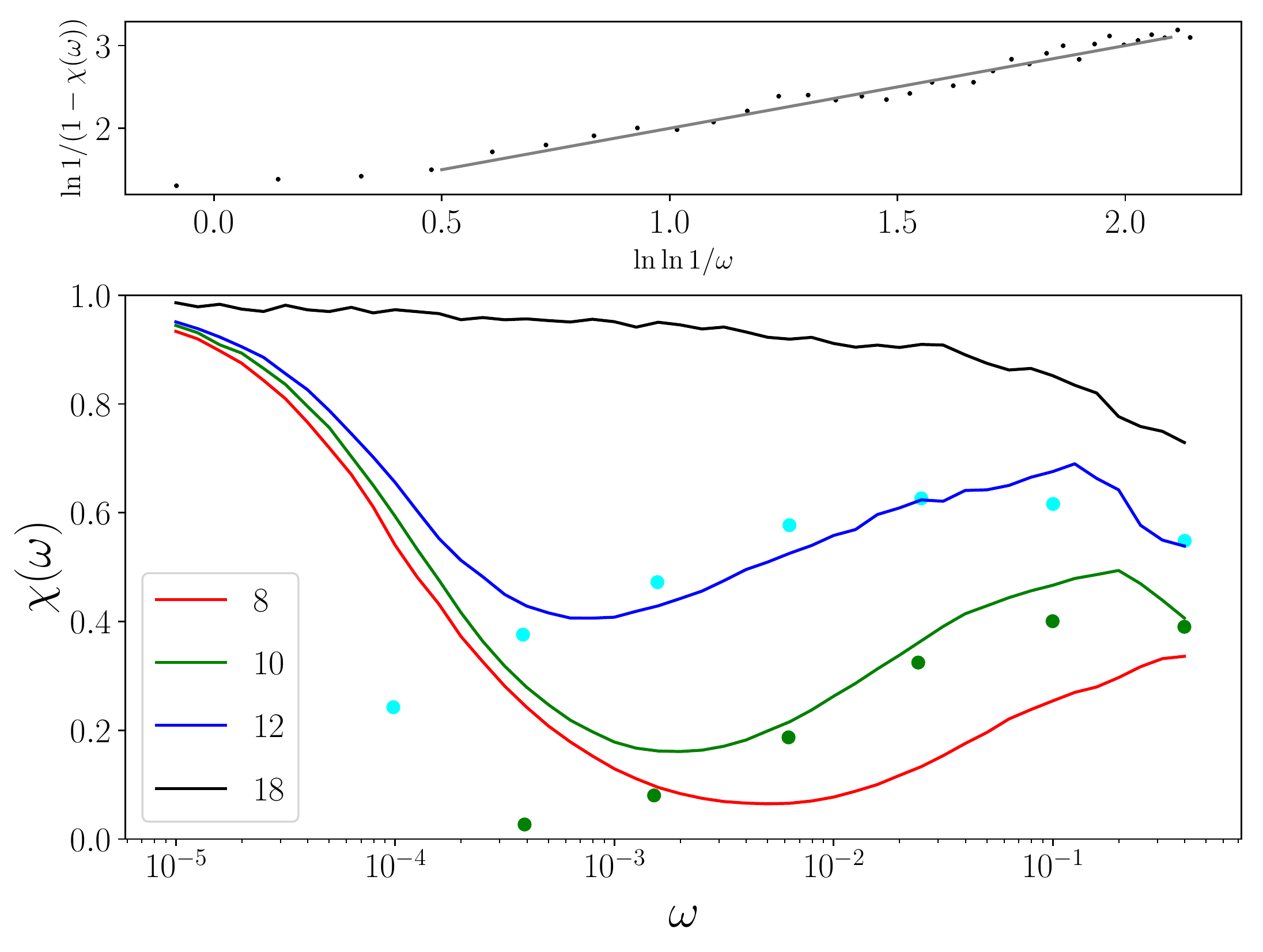}
\caption{Relative level number variance $\chi(\omega)$ as obtained by exact diagonalization. System with $N=65536$ sites and several disorder values corresponding to the delocalized phase ($W=8$, 10, and 12) and the critical point ($W=18$). 
Also shown are results of Ref.~\onlinecite{metz2017level} obtained by solution of the self-consistency equation (green dots: $W=10$, cyan dots: $W=12.5$) in the limit of $N \to \infty$ at fixed $\omega$; such limiting procedure discards the RMT contribution. Upper inset: $\ln (1-\chi(\omega))^{-1}$ vs $ \ln \ln (1/\omega)$ at the critical disorder $W=18$. The straight line corresponds to Eq.~(\ref{chi-critical-fit}) with $\mu'=1$. 
}
\label{figvar2}
\end{figure}

We now turn to discussion of exact-diagonalization numerical results for the spectral correlations, which are shown in Figs.~\ref{figvar1}, \ref{figvar2}.  In the Fig. \ref{figvar1}, the relative level number variance $\chi(\omega)$ is presented on the log-log scale. The RMT and diffusive regimes are clearly seen, cf. Fig. \ref{ls}. The numerical curve agrees very well with the analytical prediction (\ref{RWDDiff}), which is a sum of the Wigner-Dyson and diffusive contributions given by Eqs.~(\ref{chi-WD}) and (\ref{chi-RRG-diff}), respectively. 

In the Fig.~\ref{figvar2}, we show (on the log-linear scale) the evolution of $\chi(\omega)$ curve with disorder increasing towards $W_c$. In the metallic domain ($W<W_c$), both RMT and diffusive regions are observed, with the border between them (the point of the minimum of $\chi$) moving towards smaller $\omega$ with increasing disorder (i.e., increasing $N_\xi$), in agreement with Eq.~(\ref{omega-c}). 
Furthermore, the maximum value of $\chi(\omega)$ on the right border of the diffusive regime moves towards the Poisson value, Eq. (\ref{chi-star}) as predicted. In other words, the critical regime starts to develop. On the other hand, it proves to be numerically very difficult to have all three regimes (RMT, diffusive, and critical) fully developed on one curve, since this requires very large system sizes. In order to numerically observe the critical regime, we have performed simulations directly  at the critical disorder  $W=18$ (black curve). 
We see that at criticality $\chi(\omega)$ indeed gradually approaches its predicted limiting value, Eq. (\ref{chi-star}), as the frequency is lowered. Furthermore, this approach is logarithmically slow, in agreement with expectations presented above. Specifically, it can be described by 
\be
\label{chi-critical-fit}
\chi(\omega) = 1 - \frac{c^{(\chi)}}{\ln^{\mu'} (1/\omega)},
\ee
with $\mu' \simeq 1$, as shown in the inset. A further analytical work is needed to verify whether $\mu' = 1$ is indeed an exact value of this index.

Finally, let us note that the relative level number variance on RRG was recently studied\cite{metz2017level} by means of the saddle-point method that yields the self-consistency equations (\ref{2tra}) and (\ref{1tra}). The analysis of Ref.~\onlinecite{metz2017level} was performed in the limit $N\to \infty$ at fixed $\omega$, i.e., it did not include the RMT part. The results of Ref.~\onlinecite{metz2017level}  based on a numerical solution of the self-consistency equations are shown in the Fig.~\ref{figvar2} by green and cyan dots for $W=10$ and $W=12.5$, correspondingly. A good agreement between these data and results of our exact-diagonalization simulations is observed for not too low frequency, $\omega > \omega_c$, as expected. For lower frequencies our results cross over to the RMT behavior, while the data of Ref.~\onlinecite{metz2017level} continue to follow the diffusive behavior, Eq. (\ref{chi-RRG-diff}). 

\section{Summary}
\label{s5}

In this paper, we have studied dynamical and spatial correlations of eigenfunctions as well as energy level correlations in the Anderson model on RRG. Our focus was on the delocalized side of the Anderson transition and on the critical point. We have employed a supersymmetric functional integral representation to perform averaging over random potential and over configurations of the graph. This approach has allowed us to evaluate the RRG correlators  in terms of the saddle-point of the aforementioned supersymmetric action. The corresponding saddle-point equation is equivalent to the self-consistency equation characterizing an infinite Bethe lattice, and we have used analytical properties of its solution.  In addition, to obtain accurate quantitative descriptions of RRG correlation functions, we have solved the saddle-point equation numerically. Furthermore, we have studied the correlation functions by direct numerical diagonalization of the RRG problem. 
We have found an outstanding agreement between the exact diagonalization and the saddle-point approach, which serves as an additional demonstration of validity of the  latter one (and thus of the ergodicity of the delocalized phase whose analytical derivation is based on the saddle-point analysis). 

Our key findings for specific correlation functions on RRG are as follows:

\begin{itemize}

\item
We have calculated the auto-correlations of a wave function, $\alpha(r)=\left<\psi^2(r)\psi^2(0)\right>$.  The analytical results are given by Eq.~(\ref{alphadeloc}) for the delocalized phase with $N \gg N_\xi$ and by Eq.~(\ref{alphaloc}) for the critical regime. The scaling of IPR  $P_2 = N \alpha(0) \sim N_{\xi}/N$ 
(which is a direct extension of the analogous result for the SRM ensemble \cite{fyodorov1991localization})
proves the ergodicity of the delocalized phase, in agreement with the earlier numerical findings  \cite{tikhonov2016anderson,garcia-mata17}.
Exact diagonalization fully confirms these results and allows us to directly visualize the correlation length $\xi$, see Fig.~\ref{same}. There is a perfect quantitative agreement between the value of the IPR in the delocalized phase as obtained by exact diagonalization with that found from the analytical approach complemented by a numerical solution of the self-consistency equation.

\item
Next, we have studied the correlation function of different wavefunctions at the same point $\beta(0,\omega)$. 
We have found that, in the delocalized phase, it is characterized by a frequency scale $\omega\sim 1/N_{\xi}$, below which it is $\omega$-independent and is equal to $1/3$ of the autocorrelation function of a single wavefunction, $\beta(0,0)=\frac13\alpha(0)$. At a larger frequency, $\beta(0,\omega)$ crosses over to the critical behavior, see Eq.~(\ref{betascres}) and Fig. \ref{different}. Again, we have found a perfect consistency between the analytical approach (also in combination with numerical analysis of the self-consistency equation) and the exact diagonalization. 

\item
We have further extended the analysis of the correlation function $\beta(r,\omega)$ of different eigenfunctions to different points, $r \ne 0$, see Eq. (\ref{betaresr}) and Fig. \ref{sigmar}.

Interestingly, the behavior of the correlation functions $\alpha(r)$ and $\beta(r,\omega)$ can be largely understood if RRG is considered as the $d\to\infty$ limit of a $d$-dimensional lattice. This limit has, however, a very singular character in the following sense. 
First, one has to  replace $r^d\to m^r$, with $r$ staying either for distance between the points or the frequency-related scale at criticality, $L_\omega \sim \ln (1/\omega)$, or the correlation length $\xi$. Second, such a replacement misses additional subleading factors, of power-law character in $r$ or logarithmic in $\omega$. These factors are intimately related to finiteness of IPR at criticality and underlie the logarithmically slow critical dynamics. 

\item
As a quantity closely related to $\beta(0,\omega)$, we have studied the return probability $p(t)$ of a particle propagating from a given point on a RRG. In the delocalized phase, the return probability decays exponentially fast to its limiting value $N_{\xi}/N$. On the other hand, at criticality $p(t)$ decays logarithmically slowly to  a value of order unity determined by the IPR. These properties of the delocalized phase and the critical point lead to a very peculiar crossover of $p(t)$ as a function of time near the criticality: from logarithmically slow to exponentially fast variation.

\item
Finally, we have studied the level statistics on RRG, with a particular focus on the level number variance. In the delocalized phase and $N \gg N_\xi$, the statistics at relatively low frequencies $\omega$ has the RMT form, reflecting the ergodicity of the system. With increasing $\omega$, a crossover to a ``diffusive'' regime takes place, for which the relative number variance $\chi(\omega)$ and spectral compressibility $\kappa(\omega)$ increase linearly with frequency, Eq.~(\ref{chi-RRG-diff}). As a result, a new energy scale $\omega_c$  emerges, Eq.~(\ref{omega-c}), at which the ``ergodization'' of the level statistics takes place.
At criticality, the function $\chi(\omega)$ logarithmically slowly approaches its Poisson value (unity), see Eq. (\ref{chi-critical-fit}).

\end{itemize}

Let us now speculate on possible implications of our findings for many-body problems. 
As has been mentioned in the Introduction, the RRG model has important connections to many-body models studied in context of MBL. 
Indeed, many-body Hamiltonians have typically a sparse hierarchical structure in the Fock space, which establishes a link between them and RRG. We expect that the following features of the RRG problem are likely to be relevant for description of many-body eigenstates around the MBL transition: (i) quasi-localized character (Poisson level statistics) of the critical point, (ii) logarithmically slow relaxation at criticality, (iii) ergodicity of wavefunctions at the delocalized side, and (iv) ergodicity of the eigenenergy distribution in the delocalized phase characterized by the RMT level statistics at relatively low frequencies, with a crossover to $\Sigma_2(\omega)\propto\omega^2$ at higher frequency.

Indeed, all features listed above are observed in numerical studies of the MBL transition in various systems, and some of them are observed experimentally. The critical statistics, as judged by the mean adjacent gap ratio $r$, tends to the Poissonian one\cite{PhysRevB.82.174411,luitz15} upon increase of the system size, suggesting that (i) is valid. The fact that equilibration from out-of-equilibrium state is slow with power-law exponents approaching zero at the transition appears both, in numerical simulations\cite{lev2015absence,agarwal2015anomalous,luitz15,PhysRevB.92.014208,luitz2017ergodic,doggen2018many} and experiments\cite{schreiber2015observation,luschen2017observation,bordia2017probing}, confirming (ii). Many-body wavefunctions scale ergodically with the size of the Fock space in the delocalized regime \cite{luitz15,tikhonov18}, confirming (iii).  (We note that the behavior in the localized phase is more tricky, since $P_2$ shows fractal scaling in the many-body problem \cite{luitz15,tikhonov18} contrary to $P_2\sim 1$ in the localized phase on the RRG.) Very recently, the level number variance in the many-body problem has been studied numerically\cite{cotler2017black} and analytically\cite{altland2018quantum} for the SYK model, suggesting that (iv) is also valid.

While we were preparing this manuscript for publication, a preprint appeared \cite{biroli2018} that discusses statistical properties of the RRG model. The central conclusion of the authors of Ref.~\cite{biroli2018}---the ergodicity of the RRG model in the delocalized phase at $N \gg N_c$, with the crossover scale $N_c$ (which is the same as $N_{\xi}$ of the present work) diverging exponentially at the Anderson transition point---is in agreement with the present work and with the earlier works\cite{fyodorov1991localization,tikhonov2016anderson,metz2017level,garcia-mata17} on RRG and SRM models.

\section{Acknowledgments}
We are grateful to F. L. Metz and I. P. Castillo for sharing detailed notes related to Ref. \onlinecite{metz2017level} and to G. Lemari\'e for useful discussions. This work was supported by Russian Science Foundation under Grant No. 14-42-00044. KT acknowledges support by Alexander von Humboldt Foundation.
\bibliography{rrg}

\begin{thebibliography}{104}
\bibitem [{{More accurately, the disconnected part dominates over the
  $r$-dependent term (\ref{alphadeloc}) starting from the distance $r = \xi - 3
  \ln \xi /(2 \ln m)$ which is slightly smaller than $\xi$}()}]{foot2}%
  \BibitemOpen
  {More accurately, the disconnected part dominates over the $r$-dependent term
  (\ref{alphadeloc}) starting from the distance $r = \xi - 3 \ln \xi /(2 \ln
  m)$ which is slightly smaller than $\xi$},\ \href@noop {} {}\BibitemShut
  {NoStop}%
\end{thebibliography}

\end{document}